\documentclass[
amsmath, amssymb,
aps,
prb,
reprint,
superscriptaddress,
floatfix
]{revtex4-2}
\usepackage{placeins}
\usepackage{siunitx}
\usepackage{array}
\usepackage{graphicx}
\usepackage{natbib}
\usepackage{comment}
\usepackage{xcolor}
\usepackage{tabularray}
\usepackage{titlesec}
\usepackage{gensymb}
\usepackage{float}
\usepackage[normalem]{ulem}
\usepackage[utf8]{inputenc}
\usepackage{xr}
\usepackage{indentfirst} 
\makeatletter
\newcommand*{\addFileDependency}[1]{
\typeout{(#1)}
\@addtofilelist{#1}
\IfFileExists{#1}{}{\typeout{No file #1.}}
}\makeatother
\newcommand*{\myexternaldocument}[1]{%
\externaldocument{#1}%
\addFileDependency{#1.tex}%
\addFileDependency{#1.aux}%
}
\myexternaldocument{supplement}
\definecolor{LightGray}{gray}{0.9}
\titlespacing*{\section}{0pt}{2pt plus 1 pt minus 1 pt}{2 pt plus 1pt minus 1pt}

\newcommand{\Ce}{\mbox{CeGaGe}}
\newcommand{\commented}[1]{}
\begin{document}
\title{Structural characterization of the candidate Weyl semimetal \mbox{CeGaGe}}
\author{Liam J. Scanlon}
\author{Santosh Bhusal}
\affiliation{Department of Physics and Astronomy, University of Kentucky, Lexington, KY 40506 USA}
\author{Christina M. Hoffmann}
\author{Junhong He}
\affiliation{Neutron Scattering Division, Oak Ridge National Laboratory, Oak Ridge, Tennessee 37831 USA}
\author{Sean R. Parkin}
\affiliation{Department of Chemistry, University of Kentucky, Lexington, Kentucky 40506 USA}
\author{Brennan J. Arnold}
\author{William J. Gannon}
\email{Contact author: wgannon@uky.edu}
\affiliation{Department of Physics and Astronomy, University of Kentucky, Lexington, Kentucky 40506 USA}
\date{\today}
\begin{abstract}
Weyl semimetals have a variety of intriguing physical properties, including topologically protected electronic states that coexist with conducting states.
Possible exploitation of topologically protected states in a conducting material is promising for technological applications.
Weyl semimetals that form in a noncentrosymmetric structure that also contain magnetic moments may host a variety of emergent phenomena that cannot be seen in magnetic, centrosymmetric Weyl materials.
It can be difficult to distinguish definitively between a centrosymmetric structure and one of its noncentrosymmetric subgroups with standard powder X-ray diffractometers in cases where two atoms in the compound have nearly the same atomic number, as is the case for the candidate Weyl semimetal \Ce.
In these cases, a careful single-crystal neutron diffraction experiment with high-angle reflections provides complimentary information to X-ray diffraction and definitively resolves any ambiguity between centrosymmetric and noncentrosymmetric crystal structures.
Single-crystal neutron diffraction measurements on the candidate Weyl semimetal \Ce\ confirm that its structure is noncentrosymmetric, described by space group 109 $\left(I4_1md\right)$ rather than the centrosymmetric space group 141 $\left(I4_1/amd\right)$.
There are many high-angle reflections in the data set that give clear, physically intuitive evidence that \Ce\ forms with $I4_1md$ symmetry since Bragg planes of these reflections can contain Ga with no Ge or vice versa, whereas the Bragg planes for a structure with $I4_1/amd$ symmetry would have a mix of Ga and Ge.
Further, in some crystals we have studied, there is clear evidence for a structural transition from body-centered $I4_1md$ symmetry to primitive $P4_3$ and/or $P4_1$ symmetry.   
\end{abstract}
\maketitle
\section{Introduction}
The recent appreciation for the role of topology in condensed matter physics has driven the study of materials that host protected electronic states due to their potential uses in next-generation technologies \cite{TI_colloquium_Hasan2010,ElectronicPropertiesGraphene_Castro2009,GrapheneBilayerWithTwist_Andrei2020,TopoWeyl_Yan,WeylReview_Hasan2021,WeylAndDiracSemimetals_Armitage}.
Among bulk materials that potentially host such states, Weyl semimetals also feature conducting states in their bulk, allowing for the exploitation of both protected and conventional electronic states in future applications.
For candidate Weyl materials, it is critical to have a detailed understanding of the underlying symmetries in the material, as these properties govern whether or not Weyl points in the electronic band structure can exist at all and what properties these Weyl points may posses.
For a Weyl semimetal to exist, there must be either ordered magnetic moments that break time-reversal symmetry (TRS) \cite{Co2MnGa_Belopolski_Sci_2019} or broken inversion symmetry in the crystal structure \cite{LaAlGe_Xu_SciAdv_2017}.
When combined with spin-orbit coupling, these broken symmetries can lead to degenerate bands that are topologically protected with linear dispersions that mimic Weyl particles \cite{WeylAndDiracSemimetals_Armitage,TopoWeyl_Yan,WeylReview_Hasan2021}.
Some exotic topological states, for instance lines of Weyl points in the band structure known as Kramers nodal lines, can only exist in materials with broken inversion symmetry \cite{KNLmetals_Xie_NatComm_2021,SymmEnforc_Hirschmann_PRM_2021}.

Given that the physical properties observed in experiments on correlated topological matter are intricately linked to the underlying symmetries of the material, it is absolutely critical to have an understanding of these symmetries.
For a potential Weyl material, the presence or lack of TRS can be found through measurements of thermodynamic properties to determine whether a material is in a magnetically ordered state in a certain temperature range.
However, the presence or lack of inversion symmetry can be a much more subtle detail, as it relates to the intricacies of a crystal structure, and determinations are also complicated by the elements that make up the material.

Regardless, in some instances, materials with Weyl points in their band structures possess both magnetic moments that order, breaking TRS, and noncentrosymmetric crystal structures.
In such materials, it is possible to have Weyl states above the magnetic ordering temperature and study the interplay of Weyl quasiparticles with magnetic ordering.
One such family of materials with noncentrosymmetric crystal structures that have been studied in this context are the \mbox{$RXZ$} materials where \textit{R} is a rare-earth element, $X$ is Al or Ga, and $Z$ is Si or Ge.
There are at least 12 known materials in this family \cite{NdAlSi_Gaudet_NatMat_2021,NdAlSi_Wang_PRB_2022,NdAlSi_Wang_PRB_2023,NdAlSi_Li_NatComm_2023,NdAlSi_Kumar_PRB_2023,NdAlSi_Dong_PRB_2023,CeAlGe_Cho_SSRN_2022,CeAlGe_He_SCP_2023,CeAlGe_Hodovanets_PRB_2018,CeAlGe_Hodovanets_PRB_2022,CeAlGe_Piva_PRM_2023,CeAlGe_Puphal_PRL_2020,CeAlGe_Suzuki_Sci_2019,CeAlGe_Drucker_NatComm_2023,CeAlGe_Wang_PRB_2024,CePr_AlGe_Puphal_PRM_2019,CeAlSi_Alam_PRB_2023,CeAlSi_Tzschaschel_NatComm_2024,CeAlSi_Cheng_NatComm_2024,CeAlSi_Piva_PRR_2023,CeAlSi_Sakhya_PRM_2023,CeAlSi_Sun_PRB_2021,CeAlSi_Xu_AdvQT_2021,CeAlSi_Yang_PRB_2021,GdAlSi_Laha_PRB_2024,GdAlSi_Meena_JoPCM_2024,LaAl_SiGe_Cao_PRB_2022,LaAl_SiGe_Ng_PRB_2021,LaAlGe_Xu_SciAdv_2017,LaAlGe_Kim_PRM_2024,NdAlGe_Dhital_PRB_2023,NdAlGe_Kikugawa_MDPI_2023,NdAlGe_Wang_SSC_2020,NdAlGe_Yang_PRM_2023,NdAlGe_Zhao_NJP_2022,NdAlGe_Kikugawa_PRB_2024,PrAlGe_Destraz_NPJ_2020,PrAlGe_Meng_APL_2019,PrAlGe_Sanchez_NatComm_2020,PrAlGe_Yang_NPJ_2022,PrAlGe_Shoriki_PNAS_2024,PrAlGeySix_Yang_APL_2020,PrAlSi_Lyu_PRB_2020,PrAlSi_Wu_NPJ_2023,PrSm_AlSi_Lou_PRB_2023,SmAlSi_Yao_PRX_2023,SmAlSi_Cao_CPL_2022,SmAlSi_Xu_JoPCM_2022,CeGaSi_Gong_PRB_2024,LaCePrNd_AlGe_Lu_ML_2023,PrNdSm_AlSi_Bouaziz_PRB_2024,LaCePr_AlSi_Lu_JoMS_2024}.
When the rare-earth element is La, there are no magnetic moments, and TRS is preserved \cite{LaAl_SiGe_Cao_PRB_2022,LaAl_SiGe_Ng_PRB_2021,LaAlGe_Xu_SciAdv_2017,LaAlGe_Kim_PRM_2024}.
When the rare-earth element is Ce, Pr, Nd, Sm, or Gd, the materials order magnetically and break TRS below the ordering temperature \cite{NdAlSi_Gaudet_NatMat_2021,NdAlSi_Wang_PRB_2022,NdAlSi_Wang_PRB_2023,NdAlSi_Li_NatComm_2023,NdAlSi_Kumar_PRB_2023,NdAlSi_Dong_PRB_2023,CeAlGe_Cho_SSRN_2022,CeAlGe_He_SCP_2023,CeAlGe_Hodovanets_PRB_2018,CeAlGe_Hodovanets_PRB_2022,CeAlGe_Piva_PRM_2023,CeAlGe_Puphal_PRL_2020,CeAlGe_Suzuki_Sci_2019,CeAlGe_Drucker_NatComm_2023,CeAlGe_Wang_PRB_2024,CePr_AlGe_Puphal_PRM_2019,CeAlSi_Alam_PRB_2023,CeAlSi_Tzschaschel_NatComm_2024,CeAlSi_Cheng_NatComm_2024,CeAlSi_Piva_PRR_2023,CeAlSi_Sakhya_PRM_2023,CeAlSi_Sun_PRB_2021,CeAlSi_Xu_AdvQT_2021,CeAlSi_Yang_PRB_2021,GdAlSi_Laha_PRB_2024,GdAlSi_Meena_JoPCM_2024,NdAlGe_Dhital_PRB_2023,NdAlGe_Kikugawa_MDPI_2023,NdAlGe_Wang_SSC_2020,NdAlGe_Yang_PRM_2023,NdAlGe_Zhao_NJP_2022,NdAlGe_Kikugawa_PRB_2024,PrAlGe_Destraz_NPJ_2020,PrAlGe_Meng_APL_2019,PrAlGe_Sanchez_NatComm_2020,PrAlGe_Yang_NPJ_2022,PrAlGe_Shoriki_PNAS_2024,PrAlGeySix_Yang_APL_2020,PrAlSi_Lyu_PRB_2020,PrAlSi_Wu_NPJ_2023,PrSm_AlSi_Lou_PRB_2023,SmAlSi_Yao_PRX_2023,SmAlSi_Cao_CPL_2022,SmAlSi_Xu_JoPCM_2022,CeGaSi_Gong_PRB_2024,LaCePrNd_AlGe_Lu_ML_2023,PrNdSm_AlSi_Bouaziz_PRB_2024,LaCePr_AlSi_Lu_JoMS_2024}.
As single crystals, each member of this family has tetragonal, noncentrosymmetric $I4_1md$ symmetry (space group 109).

A wide variety of physics is found in this material family.
For instance, in \mbox{NdAlSi}, the ordered magnetic structure is helical, with an incommensurate ordering vector that matches the nesting of the Weyl nodes in the band structure\commented{is directly linked to the Weyl points in the band structure} \cite{NdAlSi_Gaudet_NatMat_2021}.
In \mbox{CeAlGe}, the magnetic order is a long-wavelength meron-antimeron type order, where the topology of the order can be tuned with magnetic field while simultaneously changing the magnitude of the observed topological Hall effect \cite{CeAlGe_Puphal_PRL_2020}.
In \mbox{CeAlSi}, the positions of the Weyl nodes move in reciprocal space with the onset of magnetic order \cite{CeAlSi_Yang_PRB_2021}.
In \mbox{PrAlSi}, a magnetic field-induced Lifshitz transition has been observed \cite{PrAlSi_Wu_NPJ_2023}.

Several of these materials were first identified not as high-quality single crystals but rather in polycrystalline form.
For example, \mbox{CeAlGe} and \mbox{NdAlSi} were both first produced as polycrystals via arc-melting \cite{CeAlGe_Zhao1990,CeAlGe_Dhar1992,NdAlSi_Wei2006}.
In these instances, the materials were reported to be alloys, with symmetry given as $I4_1/amd$ (space group 141) based on powder X-ray diffraction (PXRD) measurements.

\begin{figure}
	\begin{center}
	\includegraphics[width=\linewidth]{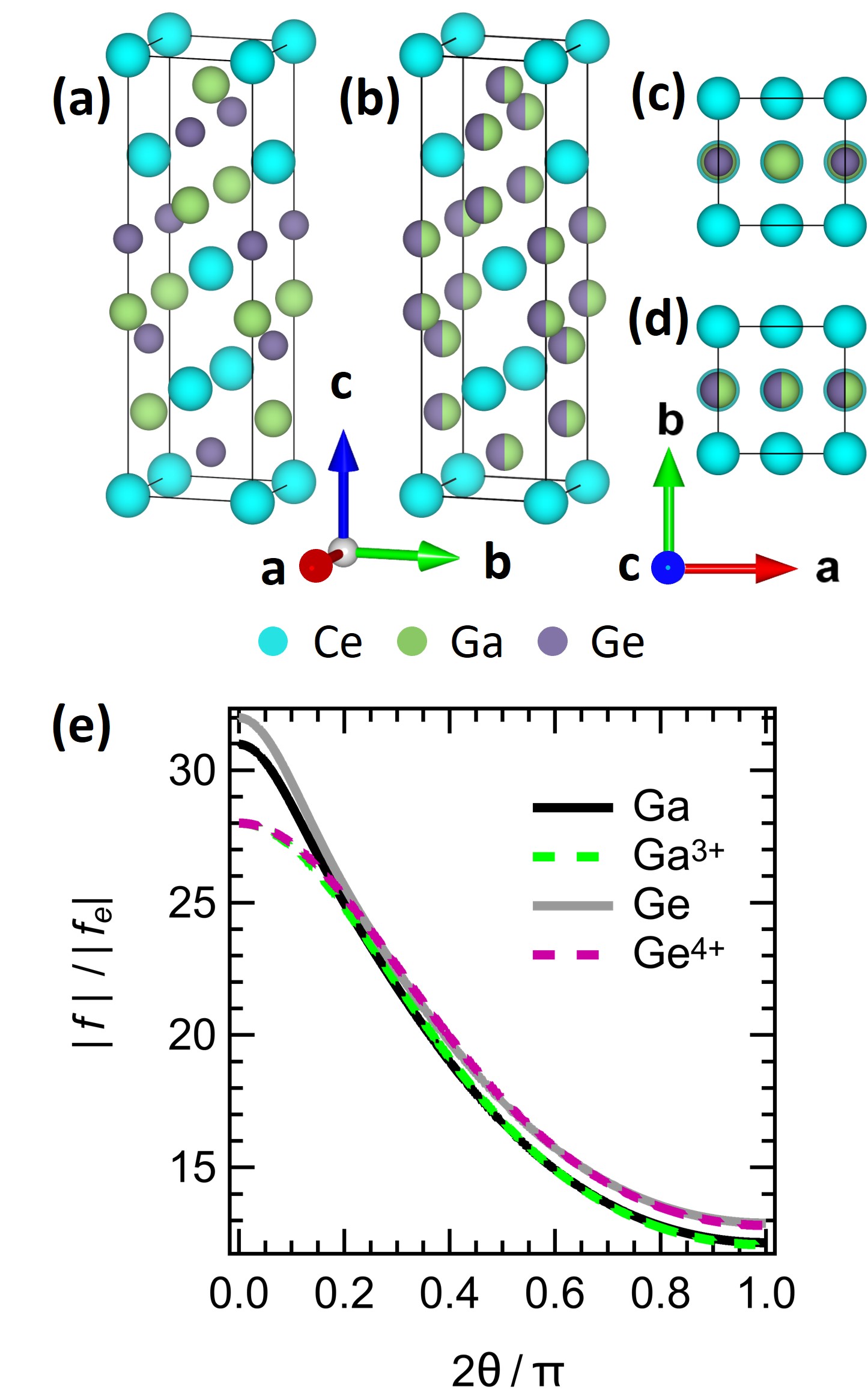}
	\end{center}
	\caption{Structure of \Ce:
    The unit cell of \Ce\ given (a) $I4_1md$ and (b) $I4_1/amd$ symmetries. Ce (cyan), Ga (green), and Ge (purple) are shown.
    For the $I4_1md$ space group, Ga and Ge occupy distinct crystallographic sites, each with Wyckoff symmetry $4a$.
    For the $I4_1/amd$ space group, Ga and Ge are shown as a random mixture on the crystallographic site with Wyckoff symmetry $8e$.
    Structures with the same (c) $I4_1md$ and (d) $I4_1/amd$ symmetries are shown projected along the crystal $c$ axis.
    (e) The normalized X-ray atomic scattering factors as a function of scattering angle $2\theta$ are shown for elemental Ga (black solid line) and Ge (gray solid line) as well as the common $\mathrm{Ga^{3+}}$ (green dashed line) and $\mathrm{Ge^{4+}}$ (magenta dashed line) oxidation states calculated using the method prescribed in Ref. \cite{IUCR_formFactor}.
    %
    }
	\label{fig:StructsAndFormFactors}
\end{figure}

The difference between $I4_1md$ and $I4_1/amd$ in the context of these materials is the positioning of the $X$ and $Z$ atoms in the structure.
With noncentrosymmetric $I4_1md$ symmetry, the $X$ and $Z$ atoms occupy distinct crystallographic sites (with Wyckoff site symmetry $4a$), shown in Figs.~\ref{fig:StructsAndFormFactors}(a) and \ref{fig:StructsAndFormFactors}(c) for the material \Ce, the subject of this paper.
With $I4_1/amd$ symmetry, the $X$ and $Z$ atoms would need to randomly occupy the same crystallographic site (with site symmetry $8e$) with 50\% occupancy, restoring inversion symmetry in the structure [Figs.~\ref{fig:StructsAndFormFactors}(b) and \ref{fig:StructsAndFormFactors}(d)].
%

In arc-melted samples, the random polycrystalline $I4_1md$ grains would be practically indistinguishable from an $I4_1/amd$ alloy in a PXRD experiment, especially due to the possibility of inversion twinning with the $I4_1md$ symmetry.
Since the band structure in a material is intimately linked to the crystal structure, it is critical to have an accurate structure determined from high-quality single crystals to fully understand the emergent physics.
In the case of \mbox{CeAlGe}, this was done with careful single-crystal X-ray diffraction (SCXRD) since the atomic scattering factors of Al and Ge are significantly different \cite{CeAlGe_Hodovanets_PRB_2018,IUCR_formFactor}.
For \mbox{NdAlSi}, a careful single-crystal neutron diffraction (SCND) experiment was performed due to the similarities in the X-ray atomic scattering factor of Al and Si \cite{NdAlSi_Gaudet_NatMat_2021,IUCR_formFactor}.

The subject of this paper \Ce\ is also a member of the $RXZ$ family.
Authors of early reports on arc-melted samples determined the crystal structure to have $I4_1/amd$ symmetry from PXRD \cite{CeGaGe_Grin, CeXX_GaGe_Pott}.
In \Ce, ferromagnetic order was seen at $T_C = 5.5~\mathrm{K}$, with a Sommerfeld coefficient $\gamma = \mathrm{40~ mJ/mol~K^2}$ \cite{CeGaGe_Dhar}, significantly larger than in \mbox{CeAlGe} ($\gamma = \mathrm{0.93~ mJ/mol~K^2}$) \cite{CeAlGe_Hodovanets_PRB_2018}, suggesting enhanced electronic correlations.
Authors of a more recent report on single crystals of \Ce\ grown from flux have confirmed the magnetic ordering at $T_C=5.5~\mathrm{K}$ but found a considerably smaller Sommerfeld coefficient $\gamma=\mathrm{13~mJ/mol~K^2}$ \cite{CeGaGe_andPrGaGe_Ram_RPB_2023}.
Authors of all of these reports have confirmed that \mbox{CeGaGe} is a poor metal, with a residual resistivity ratio as high as 2 in the likely higher-quality flux-grown crystals.

\begin{figure*}
  \setlength{\abovecaptionskip}{2mm}
  \setlength{\belowcaptionskip}{2mm}
	\begin{center}
	\includegraphics[width=\linewidth]{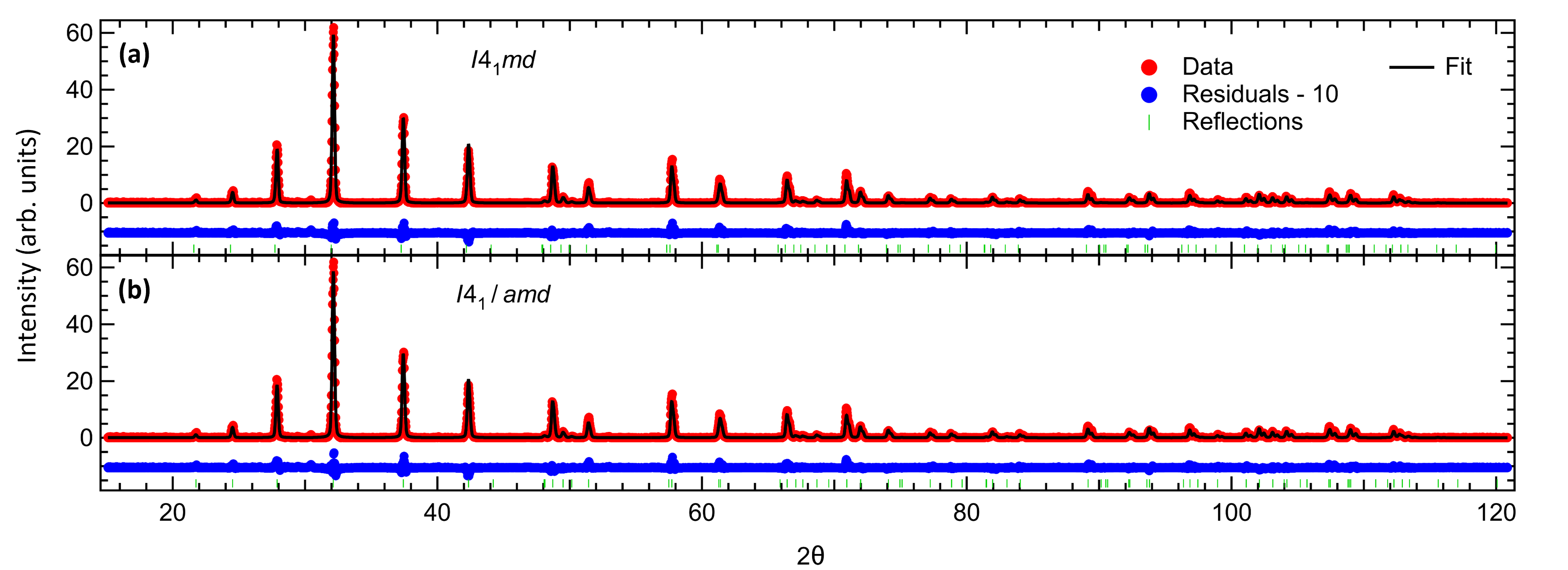}
	\end{center}
	\caption{Powder X-ray diffraction (PXRD):
    Scattered X-ray intensity as a function of scattering angle $2\theta$ measured by PXRD (red) and calculated by Rietveld refinements (black) to models with (a) $I4_1md$ and (b) $I4_1/amd$ symmetries.
    Refinements are nearly identical in both panels.
    Expected positions of Bragg reflections (green hatches) and offset residuals (blue) for each fit are shown.
    These refinements are virtually indistinguishable. 
  }
	\label{fig:pxrd}
\end{figure*}

The central problem is that there has not yet been a definitive structural determination for \Ce.
This is not possible using standard powder X-ray characterization tools available at most institutions.
As discussed above, structural determination would not have been possible in early polycrystalline samples with PXRD.
However, even a PXRD experiment on a sample made from crushed single crystals would struggle to determine the structure in the $RXZ$ materials because of the random orientation of the powder grains.
It was shown that PXRD data from flux-grown single crystals of \Ce\ fits well to a model with $I4_1md$ symmetry \cite{CeGaGe_andPrGaGe_Ram_RPB_2023}, but a side-by-side comparison of refinements to models with $I4_1md$ and $I4_1/amd$ symmetries has not been made.
PXRD data collected from a sample made from crushed single crystals is shown in Figs.~\ref{fig:pxrd}(a) and \ref{fig:pxrd}(b).
Rietveld refinements \cite{RietveldRefinement} using models with $I4_1md$ [Fig.~\ref{fig:pxrd}(a)] and $I4_1/amd$ [Fig.~\ref{fig:pxrd}(b)] symmetries are virtually identical and give equivalently good descriptions of the data.
This is discussed further in Sec. \ref{SynthesisAndCharacterization}.
In \Ce\ specifically, SCXRD is complicated by the similarities of X-ray scattering factors for Ga and Ge.
The X-ray scattering factors for Ga, Ge, and their common oxidation states are plotted in Fig.~\ref{fig:StructsAndFormFactors}(e).
There is little difference between Ga and Ge across 180\degree \ in scattering angle $2\theta$.
A single-crystal structural determination avoiding the powder grain issue that also uses a complimentary technique to increase the contrast between the Ga and Ge sites would definitively determine the structure when combined with PXRD and SCXRD measurements that can be made at many institutions.

This same issue was faced with the material \mbox{NdAlSi}, where the similarity of Al and Si in XRD measurements made structural determination difficult, even with high-quality single crystals \cite{NdAlSi_Gaudet_NatMat_2021}.
For that material, the structure was definitively determined to have $I4_1md$ symmetry using SCND.
Al and Si have significantly different coherent neutron scattering lengths \cite{NeutronScatteringLengths, NeutronScatteringLengths_URL}, making SCND a straightforward way to determine the structure.
%

We have performed a similar SCND experiment on \Ce\, as well as a number of SCXRD experiments.
We  find that the structure has $I4_1md$ symmetry in all samples.
However, we also find in some samples that there is a subtle structural transition away from the body centered $I4_1md$ to the primitive, noncentrosymmetric $P4_3$ (and/or $P4_1$) symmetry upon decreasing temperature.
These results lay a foundation for understanding the broader picture of magnetic and potential Weyl physics in \Ce\ and emphasizes that extreme caution in structural determination is needed in this family of materials.
\section{Crystal Synthesis and Characterization}\label{SynthesisAndCharacterization}

Elemental Ce (Aldrich 99.9\%), Ga (Aesar 99.99999\%), and Ge (Aldrich 99.999\%) in the molar ratio \mbox{0.86:1.0:0.79} were arc-melted into two rods \mbox{$\sim$3} inches each in length.
%
%
The melted material in each rod was cut, mixed, and remelted several times to ensure compositional homogeneity.
These polycrystalline rods were then mounted and co-aligned in a Quantum Designs two-mirror floating-zone furnace.
Floating-zone refinement was done under a 3 bar argon atmosphere to minimize loss of Ga and Ce due to their relatively high vapor pressures.
While the crystal was being grown, there was a \mbox{4 scf/h} flow of Ar through the system to minimize vapor plating to the walls of the growth chamber between the samples and mirrors.
Growth was done at a rate of \mbox{8~mm/h} and each rod was rotated in opposite directions with a frequency of 5~rpm.
This yielded a crystal of \mbox{$\sim$2.5~cm} in length and 1.7~g in mass.

%
Energy dispersive X-ray spectroscopy (EDX) was used to determine composition of the sample.
EDX spectra were taken at 38 sites on the sample using a Quanta FEG 250 environmental scanning electron microscope (SEM) equipped with an Oxford Instruments XSTREAM2 EDX detector.
Each spectrum was collected with \mbox{$\sim$500 000} photon counts.
A representative spectrum can be found in Fig. \ref{fig:representativeEDXSpectrum} in the Supplemental Material \cite{SupplementalMaterialForThisManuscript}.
Data were analyzed with the Oxford Instruments AZtec data suite using 20 keV range, 2048 channels, and pulse pile-up correction \cite{PulsePileUpCorrection}.
Carbon from the SEM mounting tape, aluminum from the interior of the SEM chamber, and any possible residual oxygen were included in fits to spectra, but not included in calculating molar compositions of the elements.
Ce, Ga, and Ge showed molar ratio \mbox{0.95:1.0:0.86}, as can be seen in \mbox{Table \ref{tab:compositions}}.
All spectra were normalized to a maximum composition of 1, with other compositions adjusted accordingly.
%
%
\begin{table}
\setlength\extrarowheight{2pt}
\sisetup{group-digits=false}
    \begin{center}
	\begin{tabular}{
	 |c|
   S[table-format = 1.2(1)]|
   S[table-format = 1.3(1)]|
   S[table-format = 1.2(1)]|
	}
		\hline
    \multicolumn{4}{|c|}{Composition of $\mathrm{Ce_{\alpha} Ga_{\beta} Ge_{\gamma}}$} \\ \hline
    { } & {$\alpha$ (Ce)} & {$\beta$ (Ga)} & {$\gamma$ (Ge)} \\ \hline
		EDX    & 0.95(2) & 1.000(9)  & 0.86(1) \\ \hline
		$I4_1md$ & 0.91(2) & 1.00(2) & 0.83(2) \\ \hline
		$I4_1/amd$ & 0.93(4) & 0.96(2) & 1.00(8) \\ \hline
	\end{tabular}
    \end{center}
	\caption{
	 Composition of CeGaGe:
      The composition of the CeGaGe floating-zone refined crystal determined by EDX and by SCND refinements to models with $I4_1md$ and $I4_1/amd$ symmetries.
      The compositions from EDX data and refinement with $I4_1md$ symmetry are within uncertainty of each other.
   }
	\label{tab:compositions}
\end{table}

To confirm that models with $I4_1md$ and $I4_1/amd$ symmetries cannot be distinguished with PXRD tools, we performed an experiment using a powder prepared from the floating-zone refined crystal.
Measurements were made with a Bruker D8 Advance diffractometer equipped with a G{\"o}bel mirror and Soller slits for beam collimation and nickel filter to suppress $K\beta$ radiation.
The results of the measurements are shown in Fig.~\ref{fig:pxrd}.
Rietveld refinements \cite{RietveldRefinement} to models with $I4_1md$ and $I4_1/amd$ symmetries using the \mbox{FULLPROF} Suite analysis package \cite{FullProf} are virtually identical.
The reported $\chi^2$ values are 1.05 for $I4_1md$ and 1.08 for $I4_1/amd$.
The reliability factor $R_{wp}$ was 18.7 for $I4_1md$ and 19.0 for $I4_1/amd$.
%
%
Details of the refined structures can be seen in \mbox{Table \ref{tab:cifStatsPXRD}} in the Supplemental Material \cite{SupplementalMaterialForThisManuscript}.

%

\section{SCND Data Acquisition and Analysis}

\begin{figure}
	\begin{center}
	\includegraphics[angle=90, width=\linewidth]{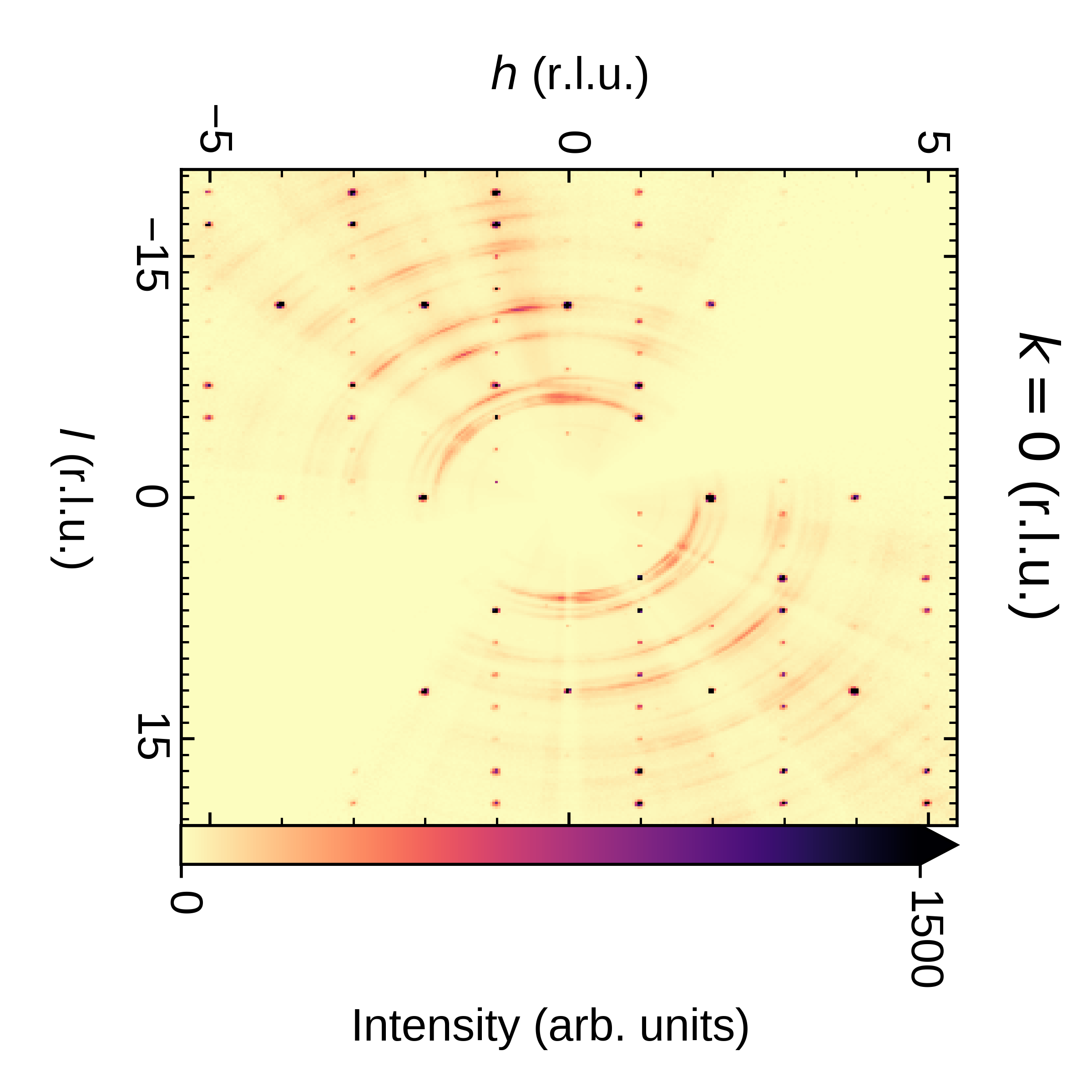}
  \end{center}
	\caption{
        The \mbox{$k=0$} plane of the merged single-crystal neutron diffraction (SCND) data before normalization.
        The data for the 19 different $\phi$ angles merged well, showing reflections only at integer indices in reciprocal space which are consistent with the reflection conditions for $I4_1md$ and $I4_1/amd$ symmetries \cite{IUCR_SpaceGroups}.
	}
	\label{fig:spectrumWideView}
\end{figure}
%
A high-resolution neutron diffraction experiment with sufficient reciprocal space coverage can accurately determine whether \Ce\ has $I4_1md$ or $I4_1/amd$ symmetry.
The bound coherent neutron scattering lengths for Ga and Ge differ by 12\% \cite{NeutronScatteringLengths,NeutronScatteringLengths_URL}, making neutron scattering quite sensitive to differences in Ga and Ge.
A single-crystal experiment eliminates the randomization of the crystal grains, while an experiment with large reciprocal space coverage can resolve features at particularly small length scales.

A SCND experiment was done using the TOPAZ diffractometer at the Spallation Neutron Source at Oak Ridge National Laboratory \cite{ORNL_instruments2018}.
A crystal with mass 3.37~mg and roughly spherical shape with radius 0.3~mm cut from our floating-zone grown crystal was chosen for the experiment.
The sample was mounted with random orientation onto an aluminum sample holder pin with glue, as is typical for TOPAZ experiments.
The sample and pin were wrapped with pure Al foil to ensure good thermal contact and loaded into a closed cycle refrigerator on the instrument and cooled to $T=100~\mathrm{K}$.
The orientation of the sample was varied by rotating around the vertical axis, and data were collected for 19 different rotation angles $\phi$.
At each angle, data were collected for $\sim$50 min\commented{ 50 $\mu\mathrm{mol}$ of neutrons} as the sample was exposed to a white beam of neutrons.
The wavelengths of the scattered neutrons were resolved by time of flight (TOF).

From these measurements, the crystal unit cell was determined to be tetragonal with \mbox{$a=b=4.2708(1)~$\AA} \ and \mbox{$c=14.5480(4)~$\AA}.
All reciprocal space scattering data are indexed in reciprocal lattice units (r.l.u).\commented{, with 1 r.l.u along the $a^*$ and $b^*$ directions equal to $2\pi / a = 1.4712~$\AA$^{-1}$ and 1 r.l.u along $c^*$ equal to $2\pi / c = 0.4319~$\AA$^{-1}$. The vectors $h$, $k$, and $l$ index scattering along the $a^*$, $b^*$, and $c^*$ directions respectively.}
A representative plot of the data in the \mbox{$k=0$} reciprocal space plane is shown in Fig.~\ref{fig:spectrumWideView}.
Measurements from all 19 rotation angles merged well and all diffraction peaks appear at integer reciprocal lattice indices, indicating a high-quality single-crystal sample.
The \mbox{$k=\pm1$} planes are shown in Fig. \ref{fig:KeqPm1DataSlices} in the Supplemental Material \cite{SupplementalMaterialForThisManuscript} to further illustrate data quality.

The raw TOF data were reduced using MANTID reduction scripts for TOPAZ measurements \cite{Mantid, MantidRelease}.
The reduction normalizes the data, accounting for absorption, neutron flux, solid angle, Lorentz factor, background, and outliers.
\begin{figure}
	\begin{center}
	\includegraphics[width=\linewidth]{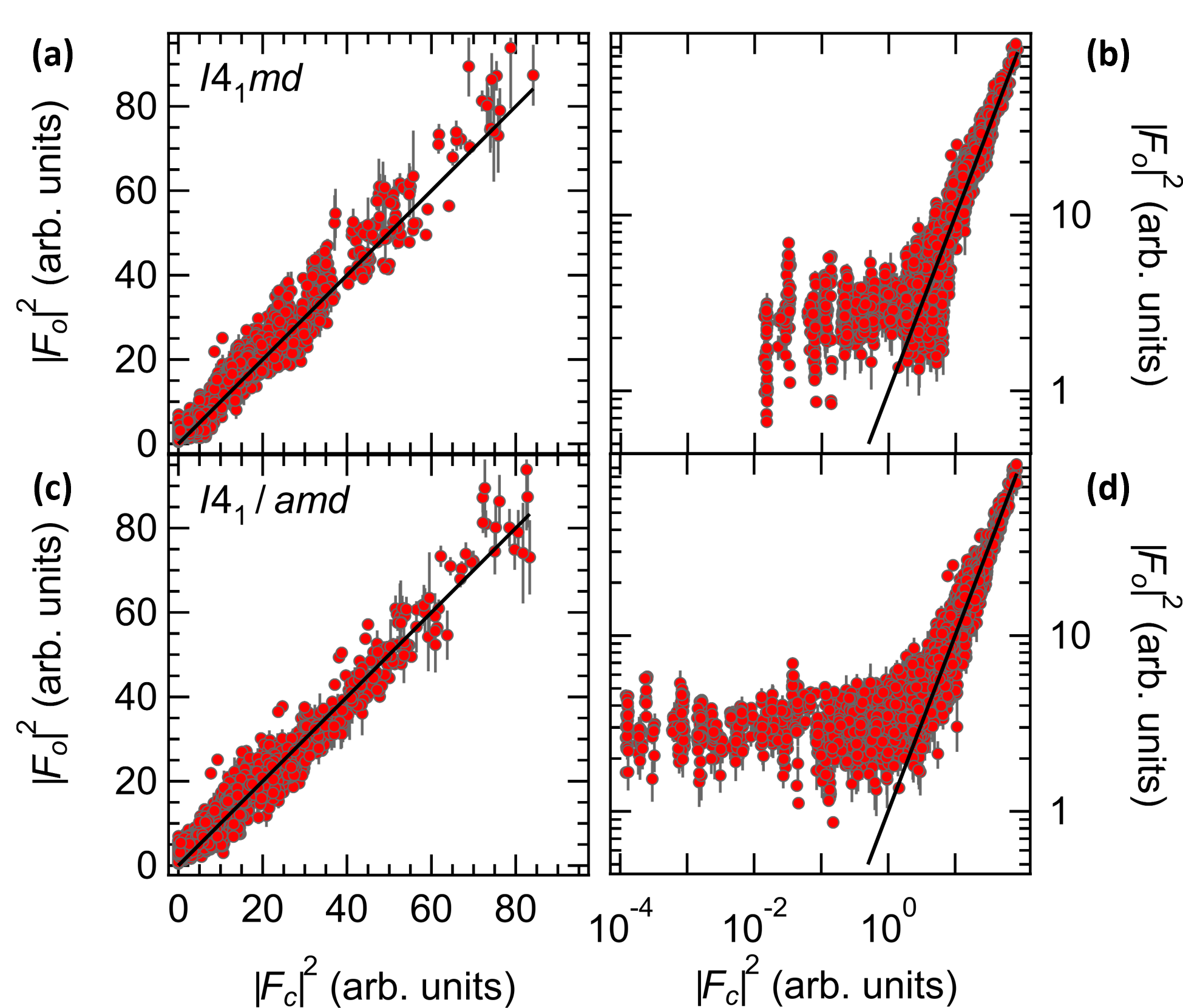}
	\end{center}
	\caption{
		Single-crystal neutron diffraction (SCND) refinements:
        Square moduli of observed vs square moduli of calculated structure factors for space groups (a) $I4_1md$ and (c) $I4_1/amd$.
        The solid black lines have slope 1 and pass through the origin.
        Plots (b) and (d) are the same plots as (a) and (c), respectively, but on a log-log scale.
	}
	\label{fig:fitplots}
\end{figure}
The reduced data give the structure factor for each observed Bragg reflection in arbitrary units.
For both structures, 4430 Bragg reflections were indexed.
Each of the structure factors had square modulus $\left|F_o\right|^2$ and uncertainty $\sigma^2_o$ such that \mbox{$\left|F_o\right|^2/\sigma^2_o\geq3$}, which is larger than the forced signal to noise cutoff of \mbox{$\left|F_o\right|^2/\sigma^2_o\geq1$}.

Structural refinement was performed using \mbox{GSAS-II} \cite{GSAS} software.
Scale, atomic position, anisotropic thermal parameters, occupancy, and\commented{secondary type 1} extinction were refined for models with $I4_1md$ and $I4_1/amd$ symmetries.
When the refinement is carried out with site exchange in a model with $I4_1md$ symmetry, there is \mbox{$<$3\%} site exchange between Ga and Ge, with uncertainty on the order of 10\%, and so the crystallographic information reported here is from a refinement that does not include site exchange.
Calculated structure factors using both the $I4_1md$ and $I4_1/amd$ space groups agree reasonably well with observed structure factors, as can be seen in Figs.~\ref{fig:fitplots}(a) and \ref{fig:fitplots}(c).
When $\left|F_o\right|^2$ is plotted as a function of the refined, calculated structure factors $\left|F_c\right|^2$, in both models, the result is a data set that is tightly grouped to a line of slope 1 passing through the origin, indicating that the refinements $\left|F_c\right|^2$ are good descriptions of our observation $\left|F_o\right|^2$.
Indeed, for the strongest reflections, there is little difference between the two crystal structures in question.

However, the reliability factors $(R)$ for the two structural models are significantly different, where $R$ is given by
\begin{equation}
	\label{eq:reliabilityFactor}
	R=\frac{\sum \left| \ \left|F_o\right|-\left|F_c\right|\ \right| }{\sum \left| F_o \right|},
\end{equation}
where $F_o$ and $F_c$ are the observed and calculated structure factors respectively.
Here, \mbox{$R=0.161$} for $I4_1md$ and \mbox{$R=0.224$} for $I4_1/amd$, a difference of 32.7\%, suggesting that $I4_1md$ is a significantly better description of the measurements.
The goodness-of-fit value of $19.13$ for $I4_1md$ is lower than that of $22.40$ for $I4_1/amd$ by 15.7\%.
The main difference in the refinements for the two structural models is seen in the weakest Bragg reflections.
When the same data from Figs.~\ref{fig:fitplots}(a) and \ref{fig:fitplots}(c) are plotted on a log-log scale to emphasize the weakest reflections [Figs.~\ref{fig:fitplots}(b) and \ref{fig:fitplots}(d)], it is clear that there are a number of peaks that are not well described by a model with $I4_1/amd$ symmetry.

This is not simply an artifact of plotting on a log-log scale.
There are 1236 data points where \mbox{$10^{-4}\leq\left|F_c\right|^2\leq 1.25$} for the $I4_1/amd$ space group.
Each of these reflections in the $I4_1/amd$ data set indexed by $hkl$ with rotation angle $\phi$, can be plotted against the corresponding reflection in the $I4_1md$ data set indexed by the same $hkl$ and $\phi$.
For 986 of the 1236 reflections, $I4_1md$ predicts more intensity than $I4_1/amd$, putting 911 of the 1236 measured $\left|F_o\right|^2$'s in better agreement with the $I4_1md$ structural model.

\begin{figure}
	\begin{center}
	\includegraphics[width=\linewidth]{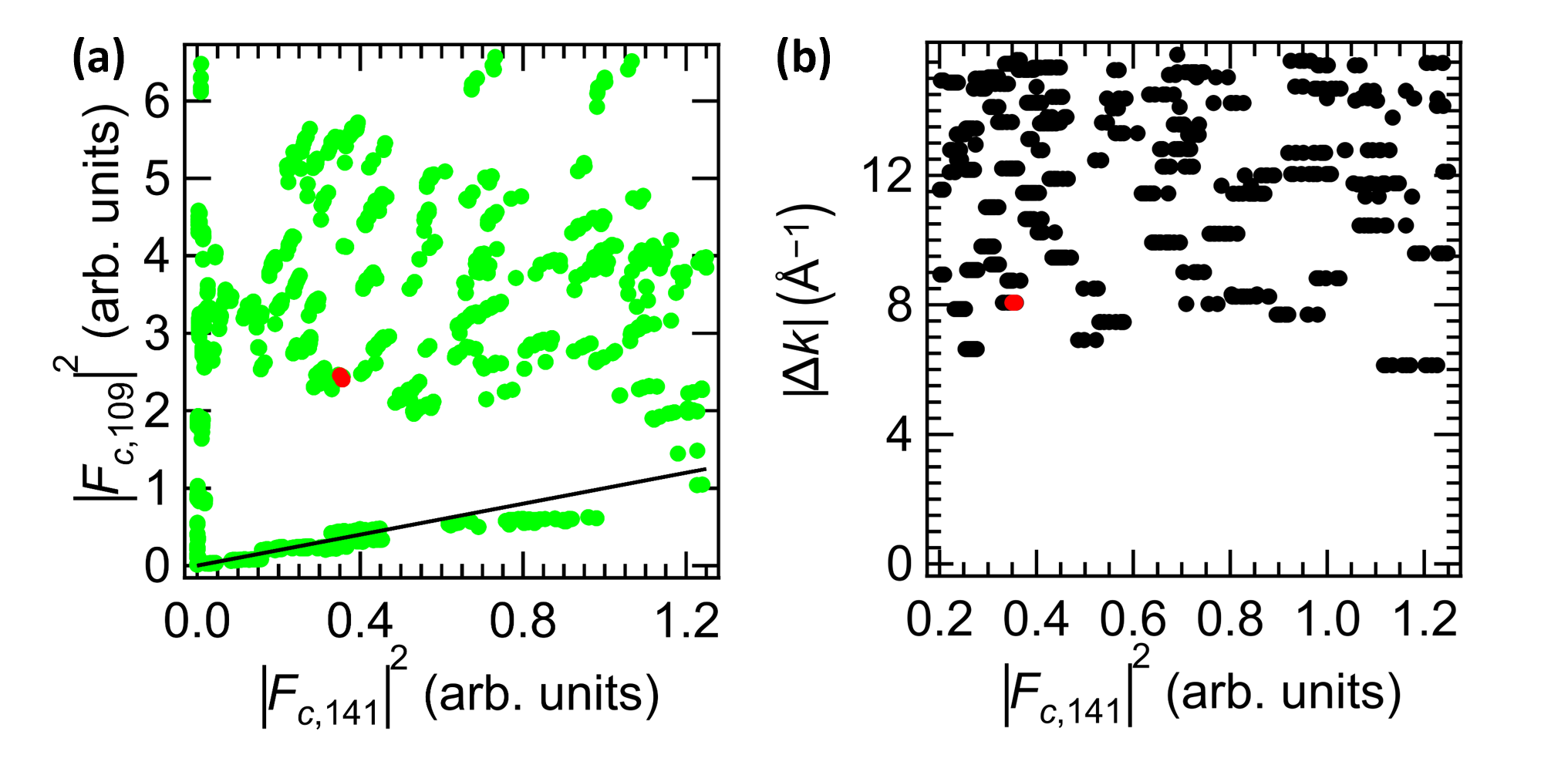}
	\end{center}
	\caption{
    Reflections with small $\left|F_c\right|^2$:
    (a) $\left|F_c\right|^2$ for the $I4_1md$ space group vs corresponding $\left|F_c\right|^2$ for $I4_1/amd$ for common Bragg reflections appearing at the same $\phi$.
    The black line has a slope of 1 and passes through the origin.
    (b) Magnitude of scattering vector as a function of $\left|F_c\right|^2$ for $I4_1/amd$.
    In both (a) and (b), points indexed by $h,k,l=-2,-2,-16$ are highlighted in red since this reflection is shown in Fig. \ref{fig:WhySg141_FoAreLower} as an example of a high quality, weak reflection.
	}
	\label{fig:Fc109vsFc141}
\end{figure}
%
The scattering vectors for these 1236 data points have relatively large magnitudes, as can be seen in Fig.~\ref{fig:Fc109vsFc141}(b).
\commented{As the length scale in reciprocal space is inversely proportional to the real space length scale being probed} Since magnitudes in reciprocal space are generally inversely proportional to magnitudes in real space, this suggests that the reflections where $I4_1md$ provides a significantly better description are representative of structure at short length scale.
\commented{Since the relevant length scale in question is 2.417 \r{A} between Ga and its nearest neighbor Ge, this is not unexpected.} Indeed, the distance between Ga and its nearest neighbor Ge is 2.417~\AA, which is a short length scale.

\begin{figure}
  \begin{center}
  \includegraphics[width=\linewidth]{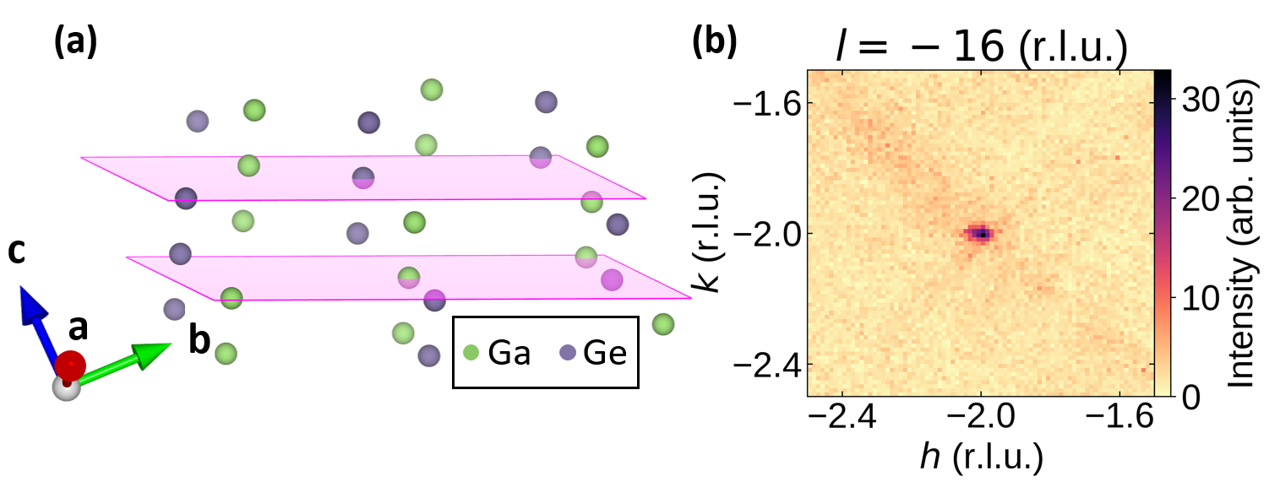}
  \end{center}
  \caption{
  (a) The crystal structure with $I4_1md$ symmetry without Ce atoms, showing Ga in green and Ge in purple.
  Lattice planes with indices $(-2,-2,-16)$, corresponding to one of the Bragg reflections where the structure factors for the $I4_1md$ and $I4_1/amd$ models differ most, are shown.
  (b) Plot of the $(-2,-2,-16)$ peak in reciprocal space, indicating that, although a weak reflection, it is clearly distinguishable from the background scattering.
  }
  \label{fig:WhySg141_FoAreLower}  
\end{figure}

Closer examination of these reflections reveals how important they are in the structural determination.
Example Bragg planes for one of the reflections in Fig.~\ref{fig:Fc109vsFc141}, the \mbox{$(-2,-2,-16)$} peak, can be seen in Fig.~\ref{fig:WhySg141_FoAreLower}(a), along with a plot of the \mbox{$(-2,-2,-16)$} peak in reciprocal space, Fig.~\ref{fig:WhySg141_FoAreLower}(b).
Although this is a weak peak, it has good signal to noise and is considerably stronger than the Al background that surrounds it.
Considering that the Bragg plane for \mbox{$(-2,-2,-16)$} can have Ga atoms with no Ge atoms or vice versa for the $I4_1md$ space group, but would include both Ga and Ge for $I4_1/amd$, it is physically intuitive why the intensities of these peaks are predicted to be significantly lower for the $I4_1/amd$ model.
A common feature of the data points in \mbox{Fig. \ref{fig:Fc109vsFc141}} is that they appear at relatively large scattering vectors and come from Bragg planes that would contain only Ga or Ge for $I4_1md$ but would contain a mix of Ga and Ge for $I4_1/amd$.

All of the refined quantities for the correct $I4_1md$ model except for the scale factors are shown in Table~\ref{tab:cifStatsAll}.
The unit cell parameters are in excellent agreement with previous reports \cite{CeGaGe_Grin,CeXX_GaGe_Pott,CeGaGe_andPrGaGe_Ram_RPB_2023}.
The average of the anisotropic thermal parameters is smaller than the isotropic thermal parameters reported in Ref. \cite{CeGaGe_Grin}.
This is expected, given that the present experiment was done at $T=100~\mathrm{K}$ while the experiment in Ref. \cite{CeGaGe_Grin} was done at room temperature.
\\

\begin{table}
\setlength\extrarowheight{2pt}
\sisetup{group-digits=false}
\begin{center}	    
  \begin{tabular}{
			|c|
      S[table-format = 1.1]|
      S[table-format = 1.1]|
      S[table-format = 1.4(1)]|
      S[table-format = 1.2(1)]|
		}
		\hline
		Atom & $x$ & $y$ & $z$ & {Occ.} \\ \hline
		Ce & 0.0 & 0.0 & 0.000 & 0.91(2) \\ \hline
		Ga & 0.0 & 0.0 & 0.4171(1) & 1.00(2) \\ \hline
		Ge & 0.0 & 0.0 & 0.5829(1) & 0.83(2) \\ \hline
	\end{tabular}
    \\
    \vspace{0.1cm}
   \begin{tabular}{
			|c|
      S[table-format = 1.4(1)]|
      S[table-format = 1.4(1)]|
      S[table-format = 1.4(1)]|
		}
		\hline
		Atom & $U_{1,1}$ & $U_{2,2}$ & $U_{3,3}$ \\ \hline
		Ce &  0.0038(8) & 0.010(1) & 0.0077(7) \\ \hline
		Ga & 0.0031(3) & 0.0146(7) & 0.0011(4) \\ \hline
		Ge & 0.019(1) & 0.019(1) & 0.013(1) \\ \hline
	\end{tabular}
  \end{center}
	\caption{
		Crystallographic information of the \Ce\ structure for $I4_1md$ refinement: The anisotropic thermal parameter $U_{1,1}$ is proportional to the mean square displacement of the atom along the $a$ axis, $U_{2,2}$ is proportional to the mean square displacement of the atom along the $b$ axis, and $U_{3,3}$ is proportional to the mean square displacement along the $c$ axis. The secondary type-1 extinction parameter $E_g$ was \mbox{$5.22\times 10^{-4}$}.\commented{As mentioned in the text, $a=b=4.2708(1)~$\AA \ and $c=14.5480(4)~$\AA.}
		}
	\label{tab:cifStatsAll}
\end{table}

\section{SCXRD}\label{sec:scxrd}

A complimentary single-crystal X-ray scattering measurement can give further structural information, especially since Friedel's law prevents observation of inversion twinning without resonant scattering, and none of the elements in \Ce\ have resonant effects in the thermal neutron range \cite{AtlasOfNeutronResonances,nistThermalNeutron}.
The only merohedric twin law allowed for $I4_1md$ is inversion twinning, which is distinguishable in a SCXRD experiment due to resonant scattering of electrons.
Moreover, the relative ease of a SCXRD experiment in a local laboratory would allow for far more rapid structural analysis of new samples grown under different conditions going forward.

A small grain of the sample used for SCND was used to conduct a SCXRD experiment.
Data were collected at \mbox{$T=100$~K} on a Bruker D8 Venture $\kappa$-axis diffractometer using molybdenum $K\alpha$ X-rays with nitrogen gas stream for cooling.
Raw data were integrated, scaled, merged, and corrected for Lorentz-polarization effects using the APEX3 package \cite{BrukerApex}.
Corrections for absorption were applied using SADABS \cite{SADABS}.
The structure was solved by iterative dual-space methods (SHELXT, \cite{SHELXT_A}), and refinement was carried out against square moduli of observed structure factors by weighted full-matrix least-squares (SHELXL, \cite{SHELXL_C}).
All atoms were refined with anisotropic displacement parameters.
Crystallographic parameters of the determined structure are summarized in Table \ref{tab:cifStatsSCXRD} in the Supplemental Material \cite{SupplementalMaterialForThisManuscript}.

The crystallographic reliability factors, defined by Eq. (\ref{eq:reliabilityFactor}), are superior for refinements using $I4_1md$ vs $I4_1/amd$ symmetry, giving \mbox{$R=0.0249$} for $I4_1md$ and \mbox{$R=0.0263$} for $I4_1/amd$. 
The goodness-of-fit values are 1.146 for $I4_1md$ and 1.258 for $I4_1/amd$.

Given all of these results, the SCXRD analysis corroborates the SCND analysis, and $I4_1md$ symmetry is the correct description for this \Ce\ sample.
Analysis of the SCXRD data with this symmetry give a Flack parameter \cite{Flack} of 0.49(8).
The small uncertainty of the Flack parameter indicates that the parameter is well defined and so the structure is noncentrosymmetric.
The Flack parameter being $\sim$0.5 indicates that there is nearly perfect inversion twinning in the sample.

\begin{figure}
        \begin{center}
        \includegraphics[width=\linewidth]{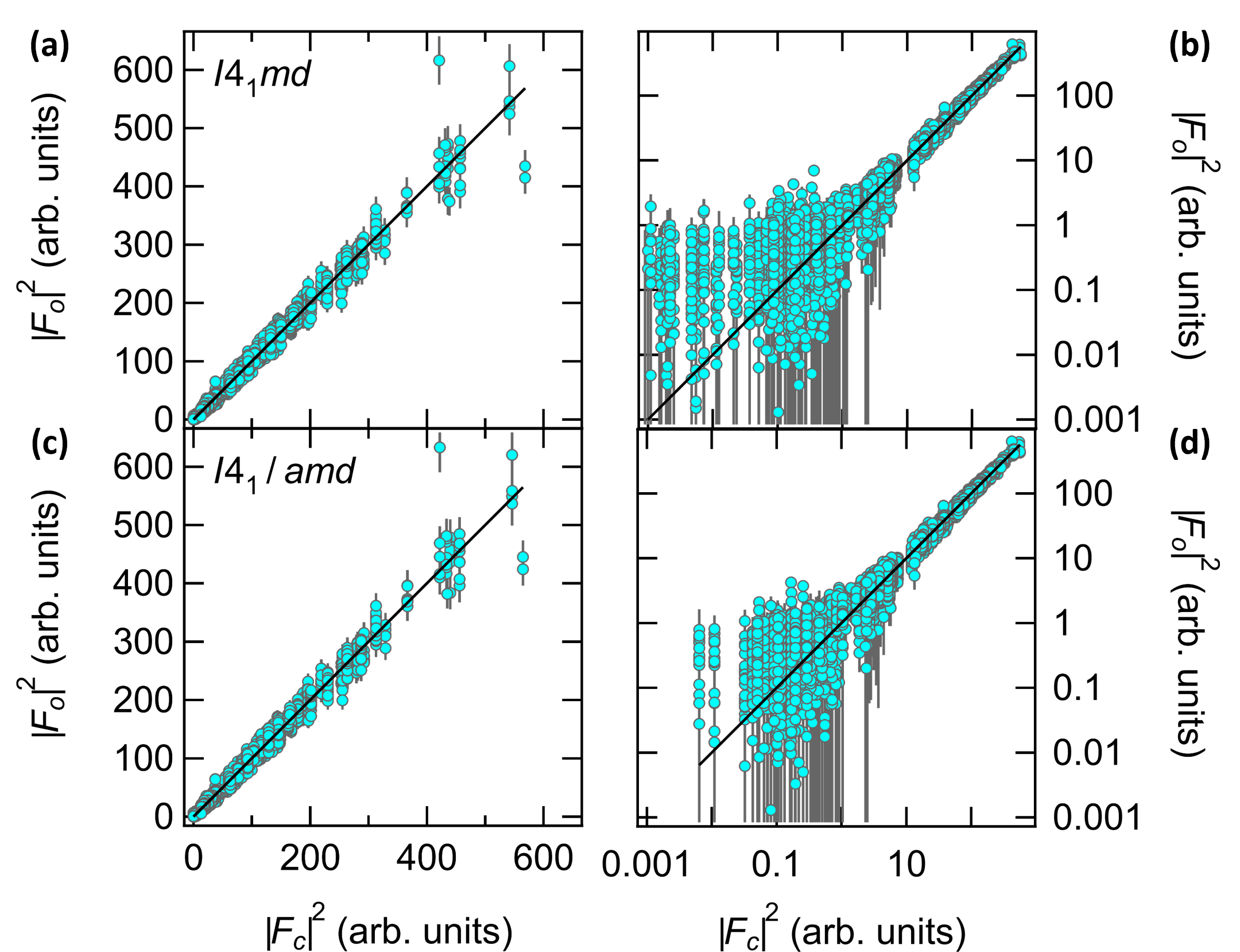}
        \caption{
           Square moduli of observed structure factors measured with single-crystal X-ray diffraction (SCXRD) vs square moduli of calculated structure factors for space groups (a) $I4_1md$ and (c) $I4_1/amd$.
           Solid black lines have slope 1 and pass through the origin.
           Plots (b) and (d) are the same plots as (a) and (c), respectively, but on a log-log scale.
           There are 181 data points where \mbox{$0.001<|F_c|^2<0.0064$} for $I4_1md$.
           For each of these 181 data points, all data points with matching $hkl$ indices for the $I4_1/amd$ data set have \mbox{$|F_c|^2=0$}.
           %
        }
        \label{fig:scxrdFo2vsFc2}
        \end{center}
    \end{figure}

These SCXRD results are plotted in Fig.~\ref{fig:scxrdFo2vsFc2} in a fashion like the SCND presentation [Fig.~\ref{fig:fitplots}].
The discrepancies between data and refinements in Figs. \ref{fig:scxrdFo2vsFc2}(b) and \ref{fig:scxrdFo2vsFc2}(d) are not nearly as pronounced as they are in the the SCND analysis, Figs. \ref{fig:fitplots}(b) and \ref{fig:fitplots}(d).
When combined, SCND and SCXRD give a more complete picture of the crystal structure of CeGaGe.
    
\section{Structural phase transition in other CeGaGe samples}\label{sec:structTransit}

Given the relative scarcity of neutron scattering time on a high-resolution diffractometer and reasonable agreement between our SCND and SCXRD results, SCXRD was used to determine the structure of additional \Ce\ crystals.
As discussed above, the crystal that was used on the TOPAZ diffractometer was found to have $I$-centered $I4_1md$ tetragonal symmetry at \mbox{$T=100$~K} in the SCND experiment and at \mbox{$T=100$~K} in the SCXRD experiment.
Additional SCXRD measurements confirm that this is the correct symmetry at room temperature for this sample.

\begin{figure}
    \begin{center}
    \includegraphics[width=\linewidth]{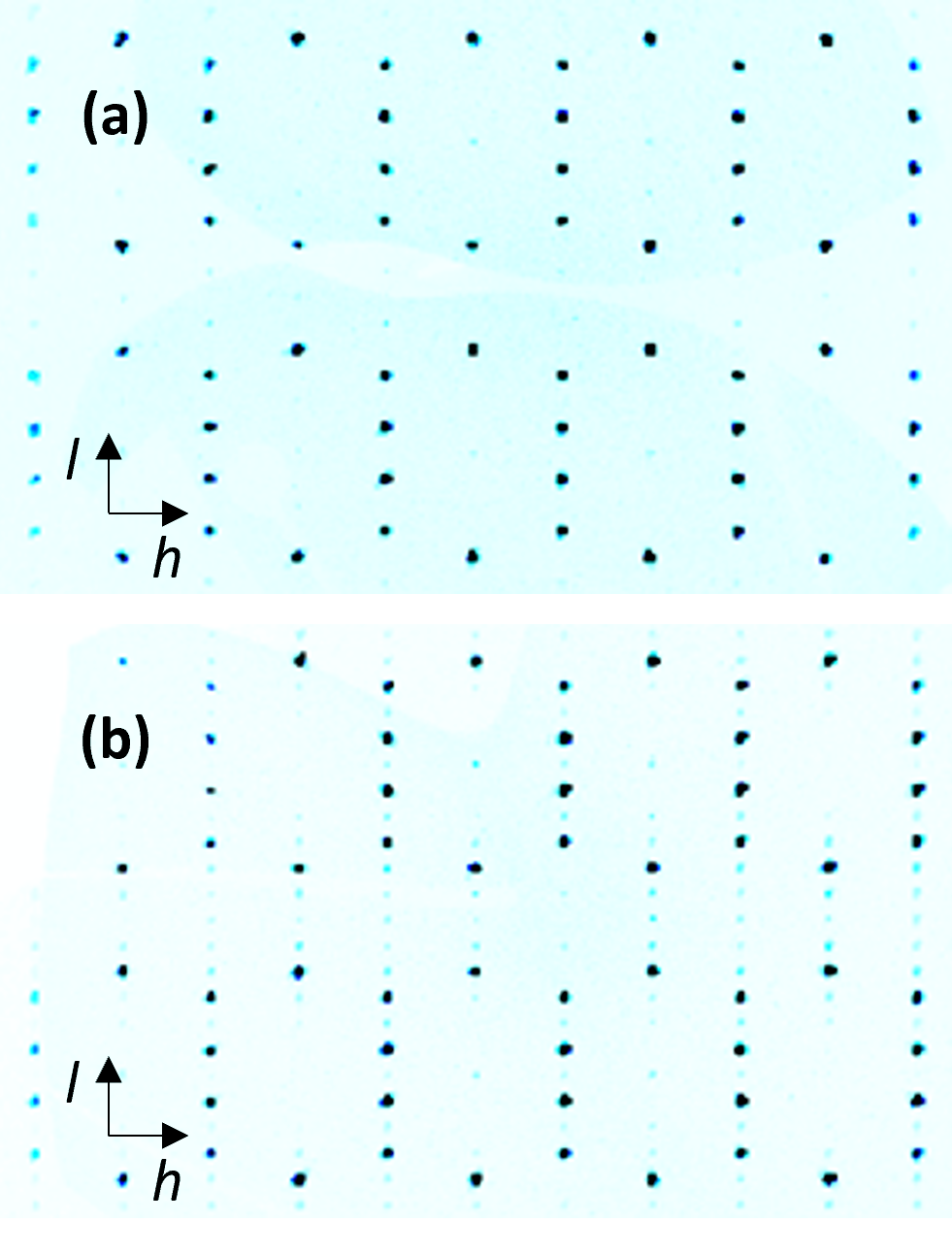}
    \end{center}
    \caption{
    Evidence of structural transition in flux-grown \Ce\ obtained with single-crystal X-ray diffraction (SCXRD).
    (a) The $k=1$ plane at room temperature.
    Intensity is on a linear scale and contrast has been enhanced to show weaker peaks more clearly.
    The $l$ axis is vertical and the $h$ axis is horizontal.
    (b) The $k=1$ plane at 100 K.
    The appearance of additional weak peaks indicates that the structure has transitioned from the body-centered tetragonal $I4_1md$ structure to the primitive tetragonal $P4_1$ or $P4_3$ structure.
    }
    \label{fig:structTransit_Flux}
\end{figure}

However, a crystal synthesized using the flux-growth technique \cite{WohlerFlux_1857,CanfieldFlux} using the same recipe as in Ref. \cite{CeGaGe_andPrGaGe_Ram_RPB_2023} shows a transition from $I$-centered to a primitive tetragonal symmetry as can be seen in Fig. \ref{fig:structTransit_Flux}.
When this sample is measured with SCXRD at room temperature, all Bragg reflections that are observed can be indexed with reflections that are expected for $I4_1md$ symmetry.
As an example, the \mbox{$k=1$} Bragg plane is shown in Fig.~\ref{fig:structTransit_Flux}(a), where only expected reflections are observed.
When the sample is cooled to \mbox{$T=100$~K} and the measurement is repeated [Fig.~\ref{fig:structTransit_Flux}(b)], weak Bragg reflections appear that are not allowed for $I4_1md$ symmetry.
These additional reflections are consistent with a transition to a primitive tetragonal structure.

SCXRD refinement shows that the structure of the flux-grown sample is well described with chiral $P4_3$ symmetry (space group 78).
A Flack parameter of 0.44(8) indicates a mix of $P4_3$ and its enantiomer $P4_1$ (space group 76) since point inversion of a chiral structure transforms it to its enantiomer~\cite{ChiralityDetInCryst_Perez2021,DetAbsConfig_Parsons_2017}.
Crystallographic information for the refinement to the $P4_3$ structure can be found in Table \ref{tab:SCXRD_P43} in the Supplemental Material \cite{SupplementalMaterialForThisManuscript}.

The difference between $I4_1md$ and $P4_3$ (or $P4_1$) is extremely subtle in this context.
Given that it occurs at a temperature between room temperature and \mbox{$T=100$~K}, the thermodynamic signature of such a minor movement of the atoms would be virtually impossible to resolve using typical laboratory-based physical property probes such as specific heat.
A more precise measurement of one of the elastic properties of the crystal would be needed to find the precise temperature of the phase transition or if the transition is continuous.  

Since we have demonstrated that it is possible to synthesize \Ce\ crystals that remain in the body-centered structure down to 100~K using float zone refining, we can speculate that the structural transition observed in the flux-grown crystals is either due to small amounts of impurities or crystallographic defects in the flux-grown samples.
Given the dependence on the physical properties of \mbox{CeAlGe} to precise stoichiometry \cite{CePr_AlGe_Puphal_PRM_2019,CeAl_SiGe_Flandorfer_JoSSC_1998,CeAlGe_Dhar1992,CeAlGe_Hodovanets_PRB_2018}, minor compositional variations could also be responsible for the symmetry change upon lowering temperature.

\section{Conclusions}
The measurements presented in this paper definitively show that the candidate Weyl semimetal \Ce\ crystallizes with noncentrosymmetric $I4_1md$ symmetry at room temperature and, in some samples, remains in this symmetry at cryogenic temperatures.
For those samples that show a structural transition, it is a transition to another noncentrosymmetric symmetry.
This structural transition is extremely subtle.
We do not observe it in measurements of the bulk thermodynamic properties, as it involves atomic displacements that are below the level expected from thermal contractions at these temperatures.
A careful measurement of the temperature dependence of the elastic constants would be required to resolve the temperature of the phase transition.

With any quantum material, having high-quality samples is extremely important, but since the electronic band structure is intimately linked to the crystal structure and its symmetries, having high-quality, well-characterized samples is especially important for materials thought to host topologically protected states.
As we emphasize in this paper, in the $RXZ$ materials, a simple powder diffraction experiment is not enough.
These experiments do not resolve differences between centrosymmetric and noncentrosymmetric structures.
In \Ce, we clearly see a temperature dependent change in crystal symmetry.


Given its compositional and structural similarities to other materials that host an interplay between exotic electronic states and magnetism, \Ce\ is an excellent candidate for study in this context.
The strong magnetic anisotropy and possible signatures of multiple magnetic phases seen in measurements of the magnetic susceptibility in flux-grown crystals \cite{CeGaGe_andPrGaGe_Ram_RPB_2023} suggest that it is likely that there will be rich magnetic phenomena in \Ce, which can now be understood definitively with respect to the crystal structure.
Given the strong sensitivity of the physical properties to precise stoichiometry in related \mbox{CeAlGe} \cite{CePr_AlGe_Puphal_PRM_2019} and the structural transitions discussed in Sec. \ref{sec:structTransit}, caution should be used in interpreting \Ce\ measurements.
However, now that the structure is known, any future measurements and calculations of electronic structure are on much firmer footing, both in the broad context of magnetic Weyl semimetals and in the narrower context of \Ce\ specifically.

\section*{Acknowledgments}
W.J.G. would like to acknowledge helpful conversations with C. Huang and J. Brill about these experiments.
The D8 Venture diffractometer was funded by the National Science Foundation Major Research Instrumentation Award (NSF-CHE-1625732) and by the University of Kentucky.

%
\FloatBarrier
\bibliographystyle{unsrtnat}
\bibliography{main}

\begin{thebibliography}{92}
\providecommand{\natexlab}[1]{#1}
\providecommand{\url}[1]{\texttt{#1}}
\expandafter\ifx\csname urlstyle\endcsname\relax
  \providecommand{\doi}[1]{doi: #1}\else
  \providecommand{\doi}{doi: \begingroup \urlstyle{rm}\Url}\fi

\bibitem[Hasan and Kane(2010)]{TI_colloquium_Hasan2010}
M.~Z. Hasan and C.~L. Kane.
\newblock Colloquium: Topological insulators.
\newblock \emph{Rev. Mod. Phys.}, 82:\penalty0 3045--3067, Nov 2010.
\newblock \doi{10.1103/RevModPhys.82.3045}.
\newblock URL \url{https://link.aps.org/doi/10.1103/RevModPhys.82.3045}.

\bibitem[Castro~Neto et~al.(2009)Castro~Neto, Guinea, Peres, Novoselov, and Geim]{ElectronicPropertiesGraphene_Castro2009}
A.~H. Castro~Neto, F.~Guinea, N.~M.~R. Peres, K.~S. Novoselov, and A.~K. Geim.
\newblock The electronic properties of graphene.
\newblock \emph{Rev. Mod. Phys.}, 81:\penalty0 109--162, Jan 2009.
\newblock \doi{10.1103/RevModPhys.81.109}.
\newblock URL \url{https://link.aps.org/doi/10.1103/RevModPhys.81.109}.

\bibitem[Andrei and MacDonald(2020)]{GrapheneBilayerWithTwist_Andrei2020}
Eva~Y. Andrei and Allan~H. MacDonald.
\newblock Graphene bilayers with a twist.
\newblock \emph{Nature Materials}, 19\penalty0 (12):\penalty0 1265–1275, November 2020.
\newblock ISSN 1476-4660.
\newblock \doi{10.1038/s41563-020-00840-0}.
\newblock URL \url{http://dx.doi.org/10.1038/s41563-020-00840-0}.

\bibitem[Yan and Felser(2017)]{TopoWeyl_Yan}
Binghai Yan and Claudia Felser.
\newblock Topological materials: {W}eyl semimetals.
\newblock \emph{Annual Review of Condensed Matter Physics}, 8\penalty0 (1):\penalty0 337–354, March 2017.
\newblock ISSN 1947-5462.
\newblock \doi{10.1146/annurev-conmatphys-031016-025458}.
\newblock URL \url{http://dx.doi.org/10.1146/annurev-conmatphys-031016-025458}.

\bibitem[Hasan et~al.(2021)Hasan, Chang, Belopolski, Bian, Xu, and Yin]{WeylReview_Hasan2021}
M.~Zahid Hasan, Guoqing Chang, Ilya Belopolski, Guang Bian, Su-Yang Xu, and Jia-Xin Yin.
\newblock {W}eyl, {D}irac and high-fold chiral fermions in topological quantum matter.
\newblock \emph{Nature Reviews Materials}, 6\penalty0 (9):\penalty0 784–803, April 2021.
\newblock ISSN 2058-8437.
\newblock \doi{10.1038/s41578-021-00301-3}.
\newblock URL \url{http://dx.doi.org/10.1038/s41578-021-00301-3}.

\bibitem[Armitage et~al.(2018)Armitage, Mele, and Vishwanath]{WeylAndDiracSemimetals_Armitage}
N.~P. Armitage, E.~J. Mele, and Ashvin Vishwanath.
\newblock {W}eyl and {D}irac semimetals in three-dimensional solids.
\newblock \emph{Rev. Mod. Phys.}, 90:\penalty0 015001, Jan 2018.
\newblock \doi{10.1103/RevModPhys.90.015001}.
\newblock URL \url{https://link.aps.org/doi/10.1103/RevModPhys.90.015001}.

\bibitem[Belopolski et~al.(2019)Belopolski, Manna, Sanchez, Chang, Ernst, Yin, Zhang, Cochran, Shumiya, Zheng, Singh, Bian, Multer, Litskevich, Zhou, Huang, Wang, Chang, Xu, Bansil, Felser, Lin, and Hasan]{Co2MnGa_Belopolski_Sci_2019}
Ilya Belopolski, Kaustuv Manna, Daniel~S. Sanchez, Guoqing Chang, Benedikt Ernst, Jiaxin Yin, Songtian~S. Zhang, Tyler Cochran, Nana Shumiya, Hao Zheng, Bahadur Singh, Guang Bian, Daniel Multer, Maksim Litskevich, Xiaoting Zhou, Shin-Ming Huang, Baokai Wang, Tay-Rong Chang, Su-Yang Xu, Arun Bansil, Claudia Felser, Hsin Lin, and M.~Zahid Hasan.
\newblock Discovery of topological {W}eyl fermion lines and drumhead surface states in a room temperature magnet.
\newblock \emph{Science}, 365\penalty0 (6459):\penalty0 1278–1281, September 2019.
\newblock ISSN 1095-9203.
\newblock \doi{10.1126/science.aav2327}.
\newblock URL \url{http://dx.doi.org/10.1126/science.aav2327}.

\bibitem[Xu et~al.(2017)Xu, Alidoust, Chang, Lu, Singh, Belopolski, Sanchez, Zhang, Bian, Zheng, Husanu, Bian, Huang, Hsu, Chang, Jeng, Bansil, Neupert, Strocov, Lin, Jia, and Hasan]{LaAlGe_Xu_SciAdv_2017}
Su-Yang Xu, Nasser Alidoust, Guoqing Chang, Hong Lu, Bahadur Singh, Ilya Belopolski, Daniel~S. Sanchez, Xiao Zhang, Guang Bian, Hao Zheng, Marious-Adrian Husanu, Yi~Bian, Shin-Ming Huang, Chuang-Han Hsu, Tay-Rong Chang, Horng-Tay Jeng, Arun Bansil, Titus Neupert, Vladimir~N. Strocov, Hsin Lin, Shuang Jia, and M.~Zahid Hasan.
\newblock Discovery of {L}orentz-violating type {II} {W}eyl fermions in {LaAlGe}.
\newblock \emph{Science Advances}, 3\penalty0 (6):\penalty0 e1603266, 2017.
\newblock \doi{10.1126/sciadv.1603266}.
\newblock URL \url{https://www.science.org/doi/abs/10.1126/sciadv.1603266}.

\bibitem[Xie et~al.(2021)Xie, Gao, Xu, Zhang, Hu, Gao, and Law]{KNLmetals_Xie_NatComm_2021}
Ying-Ming Xie, Xue-Jian Gao, Xiao~Yan Xu, Cheng-Ping Zhang, Jin-Xin Hu, Jason~Z. Gao, and K.~T. Law.
\newblock {K}ramers nodal line metals.
\newblock \emph{Nature Communications}, 12\penalty0 (1), May 2021.
\newblock ISSN 2041-1723.
\newblock \doi{10.1038/s41467-021-22903-9}.
\newblock URL \url{http://dx.doi.org/10.1038/s41467-021-22903-9}.

\bibitem[Hirschmann et~al.(2021)Hirschmann, Leonhardt, Kilic, Fabini, and Schnyder]{SymmEnforc_Hirschmann_PRM_2021}
Moritz~M. Hirschmann, Andreas Leonhardt, Berkay Kilic, Douglas~H. Fabini, and Andreas~P. Schnyder.
\newblock Symmetry-enforced band crossings in tetragonal materials: {D}irac and {W}eyl degeneracies on points, lines, and planes.
\newblock \emph{Phys. Rev. Mater.}, 5:\penalty0 054202, May 2021.
\newblock \doi{10.1103/PhysRevMaterials.5.054202}.
\newblock URL \url{https://link.aps.org/doi/10.1103/PhysRevMaterials.5.054202}.

\bibitem[Gaudet et~al.(2021)Gaudet, Yang, Baidya, Lu, Xu, Zhao, Rodriguez-Rivera, Hoffmann, Graf, Torchinsky, Nikoli{\'c}, Vanderbilt, Tafti, and Broholm]{NdAlSi_Gaudet_NatMat_2021}
Jonathan Gaudet, Hung-Yu Yang, Santu Baidya, Baozhu Lu, Guangyong Xu, Yang Zhao, Jose~A Rodriguez-Rivera, Christina~M Hoffmann, David~E Graf, Darius~H Torchinsky, Predrag Nikoli{\'c}, David Vanderbilt, Fazel Tafti, and Collin~L Broholm.
\newblock {W}eyl-mediated helical magnetism in {NdAlSi}.
\newblock \emph{Nat. Mater.}, 20\penalty0 (12):\penalty0 1650--1656, dec 2021.
\newblock \doi{10.1038/s41563-021-01062-8}.

\bibitem[Wang et~al.(2022)Wang, Dong, Guo, Lv, Huang, Xiang, Ren, Wang, Sun, Li, and Chen]{NdAlSi_Wang_PRB_2022}
Jin-Feng Wang, Qing-Xin Dong, Zhao-Peng Guo, Meng Lv, Yi-Fei Huang, Jun-Sen Xiang, Zhi-An Ren, Zhi-Jun Wang, Pei-Jie Sun, Gang Li, and Gen-Fu Chen.
\newblock {NdAlSi}: A magnetic {W}eyl semimetal candidate with rich magnetic phases and atypical transport properties.
\newblock \emph{Phys. Rev. B}, 105:\penalty0 144435, Apr 2022.
\newblock \doi{10.1103/PhysRevB.105.144435}.
\newblock URL \url{https://link.aps.org/doi/10.1103/PhysRevB.105.144435}.

\bibitem[Wang et~al.(2023)Wang, Dong, Huang, Wang, Guo, Wang, Ren, Li, Sun, Dai, and Chen]{NdAlSi_Wang_PRB_2023}
Jin-Feng Wang, Qing-Xin Dong, Yi-Fei Huang, Zhao-Sheng Wang, Zhao-Peng Guo, Zhi-Jun Wang, Zhi-An Ren, Gang Li, Pei-Jie Sun, Xi~Dai, and Gen-Fu Chen.
\newblock Quantum oscillations in the magnetic {W}eyl semimetal {NdAlSi} arising from strong {W}eyl fermion–4$f$ electron exchange interaction.
\newblock \emph{Physical Review B}, 108\penalty0 (2), July 2023.
\newblock ISSN 2469-9969.
\newblock \doi{10.1103/physrevb.108.024423}.
\newblock URL \url{http://dx.doi.org/10.1103/PhysRevB.108.024423}.

\bibitem[Li et~al.(2023)Li, Zhang, Wang, Liu, Guo, Rienks, Chen, Bertran, Yang, Phuyal, Fedderwitz, Thiagarajan, Dendzik, Berntsen, Shi, Xiang, and Tjernberg]{NdAlSi_Li_NatComm_2023}
Cong Li, Jianfeng Zhang, Yang Wang, Hongxiong Liu, Qinda Guo, Emile Rienks, Wanyu Chen, Francois Bertran, Huancheng Yang, Dibya Phuyal, Hanna Fedderwitz, Balasubramanian Thiagarajan, Maciej Dendzik, Magnus~H. Berntsen, Youguo Shi, Tao Xiang, and Oscar Tjernberg.
\newblock Emergence of {W}eyl fermions by ferrimagnetism in a noncentrosymmetric magnetic {W}eyl semimetal.
\newblock \emph{Nature Communications}, 14\penalty0 (1), November 2023.
\newblock ISSN 2041-1723.
\newblock \doi{10.1038/s41467-023-42996-8}.
\newblock URL \url{http://dx.doi.org/10.1038/s41467-023-42996-8}.

\bibitem[Tanwar et~al.(2023)Tanwar, Ahmad, Alam, Yao, Tafti, and Matusiak]{NdAlSi_Kumar_PRB_2023}
Pardeep~Kumar Tanwar, Mujeeb Ahmad, Md~Shahin Alam, Xiaohan Yao, Fazel Tafti, and Marcin Matusiak.
\newblock Gravitational anomaly in the ferrimagnetic topological {W}eyl semimetal {NdAlSi}.
\newblock \emph{Phys. Rev. B}, 108:\penalty0 L161106, Oct 2023.
\newblock \doi{10.1103/PhysRevB.108.L161106}.
\newblock URL \url{https://link.aps.org/doi/10.1103/PhysRevB.108.L161106}.

\bibitem[Dong et~al.(2023)Dong, Wang, Zhang, Bai, Liu, Cheng, Liu, Li, Xiang, Ren, Sun, and Chen]{NdAlSi_Dong_PRB_2023}
Qing-Xin Dong, Jin-Feng Wang, Li-Bo Zhang, Jian-Li Bai, Qiao-Yu Liu, Jing-Wen Cheng, Pin-Yu Liu, Cun-Dong Li, Jun-Sen Xiang, Zhi-An Ren, Pei-Jie Sun, and Gen-Fu Chen.
\newblock Large power factor, anomalous {N}ernst effect, and temperature-dependent thermoelectric quantum oscillations in the magnetic {W}eyl semimetal {NdAlSi}.
\newblock \emph{Phys. Rev. B}, 108:\penalty0 205143, Nov 2023.
\newblock \doi{10.1103/PhysRevB.108.205143}.
\newblock URL \url{https://link.aps.org/doi/10.1103/PhysRevB.108.205143}.

\bibitem[Cho et~al.(2022)Cho, Shon, Kim, Bae, Lee, Park, Yoon, Cho, Rawat, and Rhyee]{CeAlGe_Cho_SSRN_2022}
Keunki Cho, Won~Hyuk Shon, Kyoo Kim, Jaehan Bae, Jaewoong Lee, Chang-Soo Park, Seungha Yoon, Beongki Cho, Pooja Rawat, and Jong-Soo Rhyee.
\newblock Anisotropic metamagnetic transition and intrinsic {B}erry curvature in magnetic {W}eyl semimetal {NdAlGe}.
\newblock \emph{SSRN Electronic Journal}, 2022.
\newblock ISSN 1556-5068.
\newblock \doi{10.2139/ssrn.4217268}.
\newblock URL \url{http://dx.doi.org/10.2139/ssrn.4217268}.

\bibitem[He et~al.(2023)He, Li, Zeng, Zhu, Tan, Zhang, Cao, and Luo]{CeAlGe_He_SCP_2023}
Xiaobo He, Yuke Li, Hai Zeng, Zengwei Zhu, Shiyong Tan, Yongjun Zhang, Chao Cao, and Yongkang Luo.
\newblock Pressure-tuning domain-wall chirality in noncentrosymmetric magnetic {W}eyl semimetal {CeAlGe}.
\newblock \emph{Science China Physics, Mechanics \& Astronomy}, 66\penalty0 (3), feb 2023.
\newblock ISSN 1869-1927.
\newblock \doi{10.1007/s11433-022-2051-4}.
\newblock URL \url{http://dx.doi.org/10.1007/s11433-022-2051-4}.

\bibitem[Hodovanets et~al.(2018)Hodovanets, Eckberg, Zavalij, Kim, Lin, Zic, Campbell, Higgins, and Paglione]{CeAlGe_Hodovanets_PRB_2018}
H.~Hodovanets, C.~J. Eckberg, P.~Y. Zavalij, H.~Kim, W.-C. Lin, M.~Zic, D.~J. Campbell, J.~S. Higgins, and J.~Paglione.
\newblock Single-crystal investigation of the proposed type-{II} {W}eyl semimetal {CeAlGe}.
\newblock \emph{Phys. Rev. B}, 98:\penalty0 245132, Dec 2018.
\newblock \doi{10.1103/PhysRevB.98.245132}.
\newblock URL \url{https://link.aps.org/doi/10.1103/PhysRevB.98.245132}.

\bibitem[Hodovanets et~al.(2022)Hodovanets, Eckberg, Campbell, Eo, Zavalij, Piccoli, Metz, Kim, Higgins, and Paglione]{CeAlGe_Hodovanets_PRB_2022}
H.~Hodovanets, C.~J. Eckberg, D.~J. Campbell, Y.~Eo, P.~Y. Zavalij, P.~Piccoli, T.~Metz, H.~Kim, J.~S. Higgins, and J.~Paglione.
\newblock Anomalous symmetry breaking in the weyl semimetal {CeAlGe}.
\newblock \emph{Phys. Rev. B}, 106:\penalty0 235102, Dec 2022.
\newblock \doi{10.1103/PhysRevB.106.235102}.
\newblock URL \url{https://link.aps.org/doi/10.1103/PhysRevB.106.235102}.

\bibitem[Piva et~al.(2023{\natexlab{a}})Piva, Souza, Lombardi, Pakuszewski, Adriano, Pagliuso, and Nicklas]{CeAlGe_Piva_PRM_2023}
M.~M. Piva, J.~C. Souza, G.~A. Lombardi, K.~R. Pakuszewski, C.~Adriano, P.~G. Pagliuso, and M.~Nicklas.
\newblock Topological {H}all effect in {CeAlGe}.
\newblock \emph{Physical Review Materials}, 7\penalty0 (7), jul 2023{\natexlab{a}}.
\newblock ISSN 2475-9953.
\newblock \doi{10.1103/physrevmaterials.7.074204}.
\newblock URL \url{http://dx.doi.org/10.1103/PhysRevMaterials.7.074204}.

\bibitem[Puphal et~al.(2020)Puphal, Pomjakushin, Kanazawa, Ukleev, Gawryluk, Ma, Naamneh, Plumb, Keller, Cubitt, Pomjakushina, and White]{CeAlGe_Puphal_PRL_2020}
Pascal Puphal, Vladimir Pomjakushin, Naoya Kanazawa, Victor Ukleev, Dariusz~J. Gawryluk, Junzhang Ma, Muntaser Naamneh, Nicholas~C. Plumb, Lukas Keller, Robert Cubitt, Ekaterina Pomjakushina, and Jonathan~S. White.
\newblock Topological magnetic phase in the candidate {W}eyl semimetal {CeAlGe}.
\newblock \emph{Phys. Rev. Lett.}, 124:\penalty0 017202, Jan 2020.
\newblock \doi{10.1103/PhysRevLett.124.017202}.
\newblock URL \url{https://link.aps.org/doi/10.1103/PhysRevLett.124.017202}.

\bibitem[Suzuki et~al.(2019)Suzuki, Savary, Liu, Lynn, Balents, and Checkelsky]{CeAlGe_Suzuki_Sci_2019}
T.~Suzuki, L.~Savary, J.-P. Liu, J.~W. Lynn, L.~Balents, and J.~G. Checkelsky.
\newblock Singular angular magnetoresistance in a magnetic nodal semimetal.
\newblock \emph{Science}, 365\penalty0 (6451):\penalty0 377--381, 2019.
\newblock \doi{10.1126/science.aat0348}.
\newblock URL \url{https://www.science.org/doi/abs/10.1126/science.aat0348}.

\bibitem[Drucker et~al.(2023)Drucker, Nguyen, Han, Siriviboon, Luo, Andrejevic, Zhu, Bednik, Nguyen, Chen, Nguyen, Liu, Williams, Stone, Kolesnikov, Chi, Fernandez-Baca, Nelson, Alatas, Hogan, Puretzky, Huang, Yu, and Li]{CeAlGe_Drucker_NatComm_2023}
Nathan~C. Drucker, Thanh Nguyen, Fei Han, Phum Siriviboon, Xi~Luo, Nina Andrejevic, Ziming Zhu, Grigory Bednik, Quynh~T. Nguyen, Zhantao Chen, Linh~K. Nguyen, Tongtong Liu, Travis~J. Williams, Matthew~B. Stone, Alexander~I. Kolesnikov, Songxue Chi, Jaime Fernandez-Baca, Christie~S. Nelson, Ahmet Alatas, Tom Hogan, Alexander~A. Puretzky, Shengxi Huang, Yue Yu, and Mingda Li.
\newblock Topology stabilized fluctuations in a magnetic nodal semimetal.
\newblock \emph{Nature Communications}, 14\penalty0 (1), August 2023.
\newblock ISSN 2041-1723.
\newblock \doi{10.1038/s41467-023-40765-1}.
\newblock URL \url{http://dx.doi.org/10.1038/s41467-023-40765-1}.

\bibitem[Wang et~al.(2024)Wang, He, Lu, Zeng, Zou, Zhang, and Luo]{CeAlGe_Wang_PRB_2024}
Zhuo Wang, Xiaobo He, Fangjun Lu, Hai Zeng, Shuo Zou, Xiao-Xiao Zhang, and Yongkang Luo.
\newblock $^{27}\mathrm{Al}$ {NMR} study of the magnetic {W}eyl semimetal {CeAlGe}.
\newblock \emph{Phys. Rev. B}, 109:\penalty0 245106, Jun 2024.
\newblock \doi{10.1103/PhysRevB.109.245106}.
\newblock URL \url{https://link.aps.org/doi/10.1103/PhysRevB.109.245106}.

\bibitem[Puphal et~al.(2019)Puphal, Mielke, Kumar, Soh, Shang, Medarde, White, and Pomjakushina]{CePr_AlGe_Puphal_PRM_2019}
Pascal Puphal, Charles Mielke, Neeraj Kumar, Y.~Soh, Tian Shang, Marisa Medarde, Jonathan~S. White, and Ekaterina Pomjakushina.
\newblock Bulk single-crystal growth of the theoretically predicted magnetic weyl semimetals ${R}${AlGe} (${R}$={Pr}, {Ce}).
\newblock \emph{Physical Review Materials}, 3\penalty0 (2), February 2019.
\newblock ISSN 2475-9953.
\newblock \doi{10.1103/physrevmaterials.3.024204}.
\newblock URL \url{http://dx.doi.org/10.1103/PhysRevMaterials.3.024204}.

\bibitem[Alam et~al.(2023)Alam, Fakhredine, Ahmad, Tanwar, Yang, Tafti, Cuono, Islam, Singh, Lynnyk, Autieri, and Matusiak]{CeAlSi_Alam_PRB_2023}
Md~Shahin Alam, Amar Fakhredine, Mujeeb Ahmad, P.~K. Tanwar, Hung-Yu Yang, Fazel Tafti, Giuseppe Cuono, Rajibul Islam, Bahadur Singh, Artem Lynnyk, Carmine Autieri, and Marcin Matusiak.
\newblock Sign change of anomalous {H}all effect and anomalous {N}ernst effect in the {W}eyl semimetal {CeAlSi}.
\newblock \emph{Physical Review B}, 107\penalty0 (8), February 2023.
\newblock ISSN 2469-9969.
\newblock \doi{10.1103/physrevb.107.085102}.
\newblock URL \url{http://dx.doi.org/10.1103/PhysRevB.107.085102}.

\bibitem[Tzschaschel et~al.(2024)Tzschaschel, Qiu, Gao, Li, Guo, Yang, Zhang, Xie, Liu, Gao, Bérubé, Dinh, Ho, Fang, Huang, Nordlander, Ma, Tafti, Moll, Law, and Xu]{CeAlSi_Tzschaschel_NatComm_2024}
Christian Tzschaschel, Jian-Xiang Qiu, Xue-Jian Gao, Hou-Chen Li, Chunyu Guo, Hung-Yu Yang, Cheng-Ping Zhang, Ying-Ming Xie, Yu-Fei Liu, Anyuan Gao, Damien Bérubé, Thao Dinh, Sheng-Chin Ho, Yuqiang Fang, Fuqiang Huang, Johanna Nordlander, Qiong Ma, Fazel Tafti, Philip J.~W. Moll, Kam~Tuen Law, and Su-Yang Xu.
\newblock Nonlinear optical diode effect in a magnetic {W}eyl semimetal.
\newblock \emph{Nature Communications}, 15\penalty0 (1), April 2024.
\newblock ISSN 2041-1723.
\newblock \doi{10.1038/s41467-024-47291-8}.
\newblock URL \url{http://dx.doi.org/10.1038/s41467-024-47291-8}.

\bibitem[Cheng et~al.(2024)Cheng, Yan, Shi, Lou, Fedorov, Behnami, Yuan, Yang, Wang, Cheng, Xu, Xu, Xia, Pavlovskii, Peets, Zhao, Wan, Burkhardt, Guo, Li, Felser, Yang, and B\"{u}chner]{CeAlSi_Cheng_NatComm_2024}
Erjian Cheng, Limin Yan, Xianbiao Shi, Rui Lou, Alexander Fedorov, Mahdi Behnami, Jian Yuan, Pengtao Yang, Bosen Wang, Jin-Guang Cheng, Yuanji Xu, Yang Xu, Wei Xia, Nikolai Pavlovskii, Darren~C. Peets, Weiwei Zhao, Yimin Wan, Ulrich Burkhardt, Yanfeng Guo, Shiyan Li, Claudia Felser, Wenge Yang, and Bernd B\"{u}chner.
\newblock Tunable positions of {W}eyl nodes via magnetism and pressure in the ferromagnetic {W}eyl semimetal {CeAlSi}.
\newblock \emph{Nature Communications}, 15\penalty0 (1), February 2024.
\newblock ISSN 2041-1723.
\newblock \doi{10.1038/s41467-024-45658-5}.
\newblock URL \url{http://dx.doi.org/10.1038/s41467-024-45658-5}.

\bibitem[Piva et~al.(2023{\natexlab{b}})Piva, Souza, Brousseau-Couture, Sorn, Pakuszewski, John, Adriano, C\^oté, Pagliuso, Paramekanti, and Nicklas]{CeAlSi_Piva_PRR_2023}
M.~M. Piva, J.~C. Souza, V.~Brousseau-Couture, Sopheak Sorn, K.~R. Pakuszewski, Janas~K. John, C.~Adriano, M.~C\^oté, P.~G. Pagliuso, Arun Paramekanti, and M.~Nicklas.
\newblock Topological features in the ferromagnetic {W}eyl semimetal {CeAlSi}: Role of domain walls.
\newblock \emph{Physical Review Research}, 5\penalty0 (1), January 2023{\natexlab{b}}.
\newblock ISSN 2643-1564.
\newblock \doi{10.1103/physrevresearch.5.013068}.
\newblock URL \url{http://dx.doi.org/10.1103/PhysRevResearch.5.013068}.

\bibitem[Sakhya et~al.(2023)Sakhya, Huang, Dhakal, Gao, Regmi, Wang, Wen, He, Yao, Smith, Sprague, Gao, Singh, Lin, Xu, Tafti, Bansil, and Neupane]{CeAlSi_Sakhya_PRM_2023}
Anup~Pradhan Sakhya, Cheng-Yi Huang, Gyanendra Dhakal, Xue-Jian Gao, Sabin Regmi, Baokai Wang, Wei Wen, R.-H. He, Xiaohan Yao, Robert Smith, Milo Sprague, Shunye Gao, Bahadur Singh, Hsin Lin, Su-Yang Xu, Fazel Tafti, Arun Bansil, and Madhab Neupane.
\newblock Observation of {F}ermi arcs and {W}eyl nodes in a noncentrosymmetric magnetic {W}eyl semimetal.
\newblock \emph{Phys. Rev. Mater.}, 7:\penalty0 L051202, May 2023.
\newblock \doi{10.1103/PhysRevMaterials.7.L051202}.
\newblock URL \url{https://link.aps.org/doi/10.1103/PhysRevMaterials.7.L051202}.

\bibitem[Sun et~al.(2021)Sun, Lee, Yang, Torchinsky, Tafti, and Orenstein]{CeAlSi_Sun_PRB_2021}
Yue Sun, Changmin Lee, Hung-Yu Yang, Darius~H. Torchinsky, Fazel Tafti, and Joseph Orenstein.
\newblock Mapping domain-wall topology in the magnetic {W}eyl semimetal {CeAlSi}.
\newblock \emph{Phys. Rev. B}, 104:\penalty0 235119, Dec 2021.
\newblock \doi{10.1103/PhysRevB.104.235119}.
\newblock URL \url{https://link.aps.org/doi/10.1103/PhysRevB.104.235119}.

\bibitem[Xu et~al.(2021)Xu, Franklin, Jayakody, Yang, Tafti, and Sochnikov]{CeAlSi_Xu_AdvQT_2021}
Bochao Xu, Jacob Franklin, Amani Jayakody, Hung‐Yu Yang, Fazel Tafti, and Ilya Sochnikov.
\newblock Picoscale magnetoelasticity governs heterogeneous magnetic domains in a noncentrosymmetric ferromagnetic {W}eyl semimetal.
\newblock \emph{Advanced Quantum Technologies}, 4\penalty0 (3), February 2021.
\newblock ISSN 2511-9044.
\newblock \doi{10.1002/qute.202000101}.
\newblock URL \url{http://dx.doi.org/10.1002/qute.202000101}.

\bibitem[Yang et~al.(2021)Yang, Singh, Gaudet, Lu, Huang, Chiu, Huang, Wang, Bahrami, Xu, Franklin, Sochnikov, Graf, Xu, Zhao, Hoffman, Lin, Torchinsky, Broholm, Bansil, and Tafti]{CeAlSi_Yang_PRB_2021}
Hung-Yu Yang, Bahadur Singh, Jonathan Gaudet, Baozhu Lu, Cheng-Yi Huang, Wei-Chi Chiu, Shin-Ming Huang, Baokai Wang, Faranak Bahrami, Bochao Xu, Jacob Franklin, Ilya Sochnikov, David~E. Graf, Guangyong Xu, Yang Zhao, Christina~M. Hoffman, Hsin Lin, Darius~H. Torchinsky, Collin~L. Broholm, Arun Bansil, and Fazel Tafti.
\newblock Noncollinear ferromagnetic {W}eyl semimetal with anisotropic anomalous {H}all effect.
\newblock \emph{Phys. Rev. B}, 103:\penalty0 115143, Mar 2021.
\newblock \doi{10.1103/PhysRevB.103.115143}.
\newblock URL \url{https://link.aps.org/doi/10.1103/PhysRevB.103.115143}.

\bibitem[Laha et~al.(2024)Laha, Kundu, Aryal, Bozin, Yao, Paone, Rajapitamahuni, Vescovo, Valla, Abeykoon, Jing, Yin, Pasupathy, Liu, and Li]{GdAlSi_Laha_PRB_2024}
Antu Laha, Asish~K. Kundu, Niraj Aryal, Emil~S. Bozin, Juntao Yao, Sarah Paone, Anil Rajapitamahuni, Elio Vescovo, Tonica Valla, Milinda Abeykoon, Ran Jing, Weiguo Yin, Abhay~N. Pasupathy, Mengkun Liu, and Qiang Li.
\newblock Electronic structure and magnetic and transport properties of antiferromagnetic {W}eyl semimetal {GdAlSi}.
\newblock \emph{Phys. Rev. B}, 109:\penalty0 035120, Jan 2024.
\newblock \doi{10.1103/PhysRevB.109.035120}.
\newblock URL \url{https://link.aps.org/doi/10.1103/PhysRevB.109.035120}.

\bibitem[Meena et~al.(2024)Meena, Choudhury, Mudgal, Bagga, Tiwari, Rajput, Yadav, Malik, Maitra, and Nayak]{GdAlSi_Meena_JoPCM_2024}
Priyanka Meena, Amarjyoti Choudhury, Mohit Mudgal, Sonika Bagga, Vishnu~Kumar Tiwari, Sarita Rajput, C~S Yadav, Vivek~K Malik, Tulika Maitra, and Jayita Nayak.
\newblock Exploration of quantum oscillation in antiferromagnetic {W}eyl semimetal {GdSiAl}.
\newblock \emph{Journal of Physics: Condensed Matter}, November 2024.
\newblock ISSN 1361-648X.
\newblock \doi{10.1088/1361-648x/ad912e}.
\newblock URL \url{http://dx.doi.org/10.1088/1361-648X/ad912e}.

\bibitem[Cao et~al.(2022{\natexlab{a}})Cao, Zhao, Pei, Wang, Zhang, Ying, Zhao, Gao, Li, Yu, Gu, Chen, Liu, and Qi]{LaAl_SiGe_Cao_PRB_2022}
Weizheng Cao, Ningning Zhao, Cuiying Pei, Qi~Wang, Qinghua Zhang, Tianping Ying, Yi~Zhao, Lingling Gao, Changhua Li, Na~Yu, Lin Gu, Yulin Chen, Kai Liu, and Yanpeng Qi.
\newblock Pressure-induced superconductivity in the noncentrosymmetric {W}eyl semimetals $\mathrm{LaAl}{X}$ $({X}=\mathrm{Si},\mathrm{Ge})$.
\newblock \emph{Phys. Rev. B}, 105:\penalty0 174502, May 2022{\natexlab{a}}.
\newblock \doi{10.1103/PhysRevB.105.174502}.
\newblock URL \url{https://link.aps.org/doi/10.1103/PhysRevB.105.174502}.

\bibitem[Ng et~al.(2021)Ng, Luo, Yuan, Wu, Yang, and Shen]{LaAl_SiGe_Ng_PRB_2021}
Truman Ng, Yongzheng Luo, Jiaren Yuan, Yihong Wu, Hyunsoo Yang, and Lei Shen.
\newblock Origin and enhancement of the spin {H}all angle in the {W}eyl semimetals {LaAlSi} and {LaAlGe}.
\newblock \emph{Phys. Rev. B}, 104:\penalty0 014412, Jul 2021.
\newblock \doi{10.1103/PhysRevB.104.014412}.
\newblock URL \url{https://link.aps.org/doi/10.1103/PhysRevB.104.014412}.

\bibitem[Kim et~al.(2024)Kim, Kang, Kim, and Choi]{LaAlGe_Kim_PRM_2024}
Inseo Kim, Byungkyun Kang, Hyunsoo Kim, and Minseok Choi.
\newblock Crystallographic defects in {W}eyl semimetal {LaAlGe}.
\newblock \emph{Phys. Rev. Mater.}, 8:\penalty0 054203, May 2024.
\newblock \doi{10.1103/PhysRevMaterials.8.054203}.
\newblock URL \url{https://link.aps.org/doi/10.1103/PhysRevMaterials.8.054203}.

\bibitem[Dhital et~al.(2023)Dhital, Dally, Ruvalcaba, Gonzalez-Hernandez, Guerrero-Sanchez, Cao, Zhang, Tian, Wu, Frontzek, Karna, Meads, Wilson, Chapai, Graf, Bacsa, Jin, and DiTusa]{NdAlGe_Dhital_PRB_2023}
C.~Dhital, R.~L. Dally, R.~Ruvalcaba, R.~Gonzalez-Hernandez, J.~Guerrero-Sanchez, H.~B. Cao, Q.~Zhang, W.~Tian, Y.~Wu, M.~D. Frontzek, S.~K. Karna, A.~Meads, B.~Wilson, R.~Chapai, D.~Graf, J.~Bacsa, R.~Jin, and J.~F. DiTusa.
\newblock Multi-$k$ magnetic structure and large anomalous {Hall} effect in candidate magnetic {W}eyl semimetal {NdAlGe}.
\newblock \emph{Phys. Rev. B}, 107:\penalty0 224414, Jun 2023.
\newblock \doi{10.1103/PhysRevB.107.224414}.
\newblock URL \url{https://link.aps.org/doi/10.1103/PhysRevB.107.224414}.

\bibitem[Kikugawa et~al.(2023)Kikugawa, Terashima, Kato, Hayashi, Yamaguchi, and Uji]{NdAlGe_Kikugawa_MDPI_2023}
Naoki Kikugawa, Taichi Terashima, Takashi Kato, Momoko Hayashi, Hitoshi Yamaguchi, and Shinya Uji.
\newblock Bulk physical properties of a magnetic {W}eyl semimetal candidate {NdAlGe} grown by a laser floating-zone method.
\newblock \emph{Inorganics}, 11\penalty0 (1):\penalty0 20, January 2023.
\newblock ISSN 2304-6740.
\newblock \doi{10.3390/inorganics11010020}.
\newblock URL \url{http://dx.doi.org/10.3390/inorganics11010020}.

\bibitem[Wang et~al.(2020)Wang, Guo, Wang, and Yang]{NdAlGe_Wang_SSC_2020}
Tai Wang, Yongquan Guo, Cong Wang, and Shuowang Yang.
\newblock Correlation between non-centrosymmetic structure and magnetic properties in {W}eyl semimetal {NdAlGe}.
\newblock \emph{Solid State Communications}, 321:\penalty0 114041, November 2020.
\newblock ISSN 0038-1098.
\newblock \doi{10.1016/j.ssc.2020.114041}.
\newblock URL \url{http://dx.doi.org/10.1016/j.ssc.2020.114041}.

\bibitem[Yang et~al.(2023)Yang, Gaudet, Verma, Baidya, Bahrami, Yao, Huang, DeBeer-Schmitt, Aczel, Xu, Lin, Bansil, Singh, and Tafti]{NdAlGe_Yang_PRM_2023}
Hung-Yu Yang, Jonathan Gaudet, Rahul Verma, Santu Baidya, Faranak Bahrami, Xiaohan Yao, Cheng-Yi Huang, Lisa DeBeer-Schmitt, Adam~A. Aczel, Guangyong Xu, Hsin Lin, Arun Bansil, Bahadur Singh, and Fazel Tafti.
\newblock Stripe helical magnetism and two regimes of anomalous {H}all effect in {NdAlGe}.
\newblock \emph{Phys. Rev. Mater.}, 7:\penalty0 034202, Mar 2023.
\newblock \doi{10.1103/PhysRevMaterials.7.034202}.
\newblock URL \url{https://link.aps.org/doi/10.1103/PhysRevMaterials.7.034202}.

\bibitem[Zhao et~al.(2022)Zhao, Liu, Rahman, Meng, Ling, Xi, Tong, Bai, Tian, Zhong, Hu, Pi, Zhang, and Zhang]{NdAlGe_Zhao_NJP_2022}
Jun Zhao, Wei Liu, Azizur Rahman, Fanying Meng, Langsheng Ling, Chuanying Xi, Wei Tong, Yuming Bai, Zhaoming Tian, Yunbo Zhong, Ying Hu, Li~Pi, Lei Zhang, and Yuheng Zhang.
\newblock Field-induced tricritical phenomenon and magnetic structures in magnetic {W}eyl semimetal candidate {NdAlGe}.
\newblock \emph{New Journal of Physics}, 24\penalty0 (1):\penalty0 013010, January 2022.
\newblock ISSN 1367-2630.
\newblock \doi{10.1088/1367-2630/ac430a}.
\newblock URL \url{http://dx.doi.org/10.1088/1367-2630/ac430a}.

\bibitem[Kikugawa et~al.(2024)Kikugawa, Uji, and Terashima]{NdAlGe_Kikugawa_PRB_2024}
Naoki Kikugawa, Shinya Uji, and Taichi Terashima.
\newblock Anomalous {H}all effect in the magnetic {W}eyl semimetal {NdAlGe} with plateaus observed at low temperatures.
\newblock \emph{Phys. Rev. B}, 109:\penalty0 035143, Jan 2024.
\newblock \doi{10.1103/PhysRevB.109.035143}.
\newblock URL \url{https://link.aps.org/doi/10.1103/PhysRevB.109.035143}.

\bibitem[Destraz et~al.(2020)Destraz, Das, Tsirkin, Xu, Neupert, Chang, Schilling, Grushin, Kohlbrecher, Keller, Puphal, Pomjakushina, and White]{PrAlGe_Destraz_NPJ_2020}
Daniel Destraz, Lakshmi Das, Stepan~S. Tsirkin, Yang Xu, Titus Neupert, J.~Chang, A.~Schilling, Adolfo~G. Grushin, Joachim Kohlbrecher, Lukas Keller, Pascal Puphal, Ekaterina Pomjakushina, and Jonathan~S. White.
\newblock Magnetism and anomalous transport in the {W}eyl semimetal {PrAlGe}: possible route to axial gauge fields.
\newblock \emph{npj Quantum Materials}, 5\penalty0 (1), January 2020.
\newblock ISSN 2397-4648.
\newblock \doi{10.1038/s41535-019-0207-7}.
\newblock URL \url{http://dx.doi.org/10.1038/s41535-019-0207-7}.

\bibitem[Meng et~al.(2019)Meng, Wu, Qiu, Wang, Liu, Xia, Yuan, Chang, and Tian]{PrAlGe_Meng_APL_2019}
Biao Meng, Hao Wu, Yang Qiu, Chunlei Wang, Yong Liu, Zhengcai Xia, Songliu Yuan, Haixin Chang, and Zhaoming Tian.
\newblock Large anomalous {H}all effect in ferromagnetic {W}eyl semimetal candidate {PrAlGe}.
\newblock \emph{APL Materials}, 7\penalty0 (5):\penalty0 051110, 05 2019.
\newblock ISSN 2166-532X.
\newblock \doi{10.1063/1.5090795}.
\newblock URL \url{https://doi.org/10.1063/1.5090795}.

\bibitem[Sanchez et~al.(2020)Sanchez, Chang, Belopolski, Lu, Yin, Alidoust, Xu, Cochran, Zhang, Bian, Zhang, Liu, Ma, Bian, Lin, Xu, Jia, and Hasan]{PrAlGe_Sanchez_NatComm_2020}
Daniel~S. Sanchez, Guoqing Chang, Ilya Belopolski, Hong Lu, Jia-Xin Yin, Nasser Alidoust, Xitong Xu, Tyler~A. Cochran, Xiao Zhang, Yi~Bian, Songtian~S. Zhang, Yi-Yuan Liu, Jie Ma, Guang Bian, Hsin Lin, Su-Yang Xu, Shuang Jia, and M.~Zahid Hasan.
\newblock Observation of {W}eyl fermions in a magnetic non-centrosymmetric crystal.
\newblock \emph{Nature Communications}, 11\penalty0 (1), July 2020.
\newblock ISSN 2041-1723.
\newblock \doi{10.1038/s41467-020-16879-1}.
\newblock URL \url{http://dx.doi.org/10.1038/s41467-020-16879-1}.

\bibitem[Yang et~al.(2022)Yang, Corasaniti, Le, Yue, Hu, Hu, Petrovic, and Degiorgi]{PrAlGe_Yang_NPJ_2022}
R.~Yang, M.~Corasaniti, C.~C. Le, C.~Yue, Z.~Hu, J.~P. Hu, C.~Petrovic, and L.~Degiorgi.
\newblock Charge dynamics of a noncentrosymmetric magnetic {W}eyl semimetal.
\newblock \emph{npj Quantum Materials}, 7\penalty0 (1), October 2022.
\newblock ISSN 2397-4648.
\newblock \doi{10.1038/s41535-022-00507-w}.
\newblock URL \url{http://dx.doi.org/10.1038/s41535-022-00507-w}.

\bibitem[Shoriki et~al.(2024)Shoriki, Moriishi, Okamura, Yokoi, Usui, Murakawa, Sakai, Hanasaki, Tokura, and Takahashi]{PrAlGe_Shoriki_PNAS_2024}
Kentaro Shoriki, Keigo Moriishi, Yoshihiro Okamura, Kohei Yokoi, Hidetomo Usui, Hiroshi Murakawa, Hideaki Sakai, Noriaki Hanasaki, Yoshinori Tokura, and Youtarou Takahashi.
\newblock Large nonlinear optical magnetoelectric response in a noncentrosymmetric magnetic {W}eyl semimetal.
\newblock \emph{Proceedings of the National Academy of Sciences}, 121\penalty0 (12), March 2024.
\newblock ISSN 1091-6490.
\newblock \doi{10.1073/pnas.2316910121}.
\newblock URL \url{http://dx.doi.org/10.1073/pnas.2316910121}.

\bibitem[Yang et~al.(2020)Yang, Singh, Lu, Huang, Bahrami, Chiu, Graf, Huang, Wang, Lin, Torchinsky, Bansil, and Tafti]{PrAlGeySix_Yang_APL_2020}
Hung-Yu Yang, Bahadur Singh, Baozhu Lu, Cheng-Yi Huang, Faranak Bahrami, Wei-Chi Chiu, David Graf, Shin-Ming Huang, Baokai Wang, Hsin Lin, Darius Torchinsky, Arun Bansil, and Fazel Tafti.
\newblock Transition from intrinsic to extrinsic anomalous {H}all effect in the ferromagnetic weyl semimetal {PrAlGe\textsubscript{1-x}Si\textsubscript{x}}.
\newblock \emph{APL Materials}, 8\penalty0 (1):\penalty0 011111, 01 2020.
\newblock ISSN 2166-532X.
\newblock \doi{10.1063/1.5132958}.
\newblock URL \url{https://doi.org/10.1063/1.5132958}.

\bibitem[Lyu et~al.(2020)Lyu, Xiang, Mi, Zhao, Wang, Liu, Chen, Ren, Li, and Sun]{PrAlSi_Lyu_PRB_2020}
Meng Lyu, Junsen Xiang, Zhenyu Mi, Hengcan Zhao, Zhen Wang, Enke Liu, Genfu Chen, Zhian Ren, Gang Li, and Peijie Sun.
\newblock Nonsaturating magnetoresistance, anomalous {H}all effect, and magnetic quantum oscillations in the ferromagnetic semimetal {PrAlSi}.
\newblock \emph{Phys. Rev. B}, 102:\penalty0 085143, Aug 2020.
\newblock \doi{10.1103/PhysRevB.102.085143}.
\newblock URL \url{https://link.aps.org/doi/10.1103/PhysRevB.102.085143}.

\bibitem[Wu et~al.(2023)Wu, Chi, Zuo, Xu, Zhao, Luo, and Zhu]{PrAlSi_Wu_NPJ_2023}
Lei Wu, Shengwei Chi, Huakun Zuo, Gang Xu, Lingxiao Zhao, Yongkang Luo, and Zengwei Zhu.
\newblock Field-induced {L}ifshitz transition in the magnetic {W}eyl semimetal candidate {PrAlSi}.
\newblock \emph{npj Quantum Materials}, 8\penalty0 (1), jan 2023.
\newblock ISSN 2397-4648.
\newblock \doi{10.1038/s41535-023-00537-y}.
\newblock URL \url{http://dx.doi.org/10.1038/s41535-023-00537-y}.

\bibitem[Lou et~al.(2023)Lou, Fedorov, Zhao, Yaresko, B\"{u}chner, and Borisenko]{PrSm_AlSi_Lou_PRB_2023}
Rui Lou, Alexander Fedorov, Lingxiao Zhao, Alexander Yaresko, Bernd B\"{u}chner, and Sergey Borisenko.
\newblock Signature of weakly coupled $f$ electrons and conduction electrons in magnetic {W}eyl semimetal candidates {PrAlSi} and {SmAlSi}.
\newblock \emph{Physical Review B}, 107\penalty0 (3), January 2023.
\newblock ISSN 2469-9969.
\newblock \doi{10.1103/physrevb.107.035158}.
\newblock URL \url{http://dx.doi.org/10.1103/PhysRevB.107.035158}.

\bibitem[Yao et~al.(2023)Yao, Gaudet, Verma, Graf, Yang, Bahrami, Zhang, Aczel, Subedi, Torchinsky, Sun, Bansil, Huang, Singh, Blaha, Nikoli\'{c}, and Tafti]{SmAlSi_Yao_PRX_2023}
Xiaohan Yao, Jonathan Gaudet, Rahul Verma, David~E. Graf, Hung-Yu Yang, Faranak Bahrami, Ruiqi Zhang, Adam~A. Aczel, Sujan Subedi, Darius~H. Torchinsky, Jianwei Sun, Arun Bansil, Shin-Ming Huang, Bahadur Singh, Peter Blaha, Predrag Nikoli\'{c}, and Fazel Tafti.
\newblock Large topological {H}all effect and spiral magnetic order in the {W}eyl semimetal {SmAlSi}.
\newblock \emph{Phys. Rev. X}, 13:\penalty0 011035, Mar 2023.
\newblock \doi{10.1103/PhysRevX.13.011035}.
\newblock URL \url{https://link.aps.org/doi/10.1103/PhysRevX.13.011035}.

\bibitem[Cao et~al.(2022{\natexlab{b}})Cao, Su, Wang, Pei, Gao, Zhao, Li, Yu, Wang, Liu, Chen, Li, Li, and Qi]{SmAlSi_Cao_CPL_2022}
Weizheng Cao, Yunlong Su, Qi~Wang, Cuiying Pei, Lingling Gao, Yi~Zhao, Changhua Li, Na~Yu, Jinghui Wang, Zhongkai Liu, Yulin Chen, Gang Li, Jun Li, and Yanpeng Qi.
\newblock Quantum oscillations in noncentrosymmetric {W}eyl semimetal {SmAlSi}.
\newblock \emph{Chinese Physics Letters}, 39\penalty0 (4):\penalty0 047501, April 2022{\natexlab{b}}.
\newblock ISSN 1741-3540.
\newblock \doi{10.1088/0256-307x/39/4/047501}.
\newblock URL \url{http://dx.doi.org/10.1088/0256-307X/39/4/047501}.

\bibitem[Xu et~al.(2022)Xu, Niu, Bai, Zhu, Yuan, He, Han, Zhao, Yang, Xia, Liang, and Tian]{SmAlSi_Xu_JoPCM_2022}
Longmeng Xu, Haoyu Niu, Yuming Bai, Haipeng Zhu, Songliu Yuan, Xiong He, Yibo Han, Lingxiao Zhao, Yang Yang, Zhengcai Xia, Qifeng Liang, and Zhaoming Tian.
\newblock {S}hubnikov–de {H}aas oscillations and nontrivial topological states in {W}eyl semimetal candidate {SmAlSi}.
\newblock \emph{Journal of Physics: Condensed Matter}, 34\penalty0 (48):\penalty0 485701, October 2022.
\newblock ISSN 1361-648X.
\newblock \doi{10.1088/1361-648x/ac987a}.
\newblock URL \url{http://dx.doi.org/10.1088/1361-648X/ac987a}.

\bibitem[Gong et~al.(2024)Gong, Wang, Han, Zeng, Ma, Wang, Lin, Wang, and Xia]{CeGaSi_Gong_PRB_2024}
Jing Gong, Huan Wang, Kun Han, Xiang-Yu Zeng, Xiao-Ping Ma, Yi-Ting Wang, Jun-Fa Lin, Xiao-Yan Wang, and Tian-Long Xia.
\newblock Anomalous {H}all effect in an antiferromagnetic {CeGaSi} single crystal.
\newblock \emph{Phys. Rev. B}, 109:\penalty0 024434, Jan 2024.
\newblock \doi{10.1103/PhysRevB.109.024434}.
\newblock URL \url{https://link.aps.org/doi/10.1103/PhysRevB.109.024434}.

\bibitem[Lu et~al.(2023)Lu, Yang, Huang, Bian, Zhang, and Jia]{LaCePrNd_AlGe_Lu_ML_2023}
Hong Lu, Wentao Yang, Yuqing Huang, Yi~Bian, Xiao Zhang, and Shuang Jia.
\newblock Multi-experimental determination of magnetic transition in {W}eyl semimetals {$R$AlGe}.
\newblock \emph{Materials Letters}, 335:\penalty0 133819, March 2023.
\newblock ISSN 0167-577X.
\newblock \doi{10.1016/j.matlet.2023.133819}.
\newblock URL \url{http://dx.doi.org/10.1016/j.matlet.2023.133819}.

\bibitem[Bouaziz et~al.(2024)Bouaziz, Bihlmayer, Patrick, Staunton, and Bl\"ugel]{PrNdSm_AlSi_Bouaziz_PRB_2024}
Juba Bouaziz, Gustav Bihlmayer, Christopher~E. Patrick, Julie~B. Staunton, and Stefan Bl\"ugel.
\newblock Origin of incommensurate magnetic order in the ${R}\mathrm{AlSi}$ magnetic {W}eyl semimetals $({R}=\mathrm{Pr}, \mathrm{Nd}, \mathrm{Sm})$.
\newblock \emph{Phys. Rev. B}, 109:\penalty0 L201108, May 2024.
\newblock \doi{10.1103/PhysRevB.109.L201108}.
\newblock URL \url{https://link.aps.org/doi/10.1103/PhysRevB.109.L201108}.

\bibitem[Lu et~al.(2024)Lu, Zang, Ren, He, Rodionova, Yan, and Magomedov]{LaCePr_AlSi_Lu_JoMS_2024}
Hong Lu, Haotong Zang, Xiao Ren, Xuheng He, Valeria Rodionova, Eryun Yan, and Kurban Magomedov.
\newblock Comparative analysis of magnetoresistance in {W}eyl semimetal {$R$AlSi}.
\newblock \emph{Journal of Materials Science}, 59\penalty0 (31):\penalty0 14653–14660, July 2024.
\newblock ISSN 1573-4803.
\newblock \doi{10.1007/s10853-024-10035-6}.
\newblock URL \url{http://dx.doi.org/10.1007/s10853-024-10035-6}.

\bibitem[Zhao and Parthé(1990)]{CeAlGe_Zhao1990}
J.~T. Zhao and E.~Parthé.
\newblock Structure of {YAlGe} and isotypic rare-earth–aluminium germanides.
\newblock \emph{Acta Crystallographica Section C Crystal Structure Communications}, 46\penalty0 (12):\penalty0 2276–2279, December 1990.
\newblock ISSN 0108-2701.
\newblock \doi{10.1107/s0108270190005571}.
\newblock URL \url{http://dx.doi.org/10.1107/S0108270190005571}.

\bibitem[Dhar et~al.(1992)Dhar, Pattalwar, and Vijayaraghavan]{CeAlGe_Dhar1992}
S.K. Dhar, S.M. Pattalwar, and R.~Vijayaraghavan.
\newblock Magnetic and thermal behavior of {CeAl$X$} ({$X$}={Si} and {Ge}) compounds.
\newblock \emph{Journal of Magnetism and Magnetic Materials}, 104-107:\penalty0 1303--1304, 1992.
\newblock ISSN 0304-8853.
\newblock \doi{https://doi.org/10.1016/0304-8853(92)90593-D}.
\newblock URL \url{https://www.sciencedirect.com/science/article/pii/030488539290593D}.
\newblock Proceedings of the International Conference on Magnetism, Part II.

\bibitem[He et~al.(2006)He, Zhang, Zeng, and Zhuang]{NdAlSi_Wei2006}
W~He, J~Zhang, L~Zeng, and Y~Zhuang.
\newblock Crystal structural refinement for {NdAlSi}.
\newblock \emph{Rare Metals}, 25\penalty0 (4):\penalty0 355–358, August 2006.
\newblock ISSN 1001-0521.
\newblock \doi{10.1016/s1001-0521(06)60067-3}.
\newblock URL \url{http://dx.doi.org/10.1016/S1001-0521(06)60067-3}.

\bibitem[Brown et~al.(2006)Brown, Fox, Maslen, O’Keefe, and Willis]{IUCR_formFactor}
P.~J. Brown, A.~G. Fox, E.~N. Maslen, M.~A. O’Keefe, and B.~T.~M. Willis.
\newblock \emph{Intensity of diffracted intensities}, page 554–595.
\newblock International Union of Crystallography, October 2006.
\newblock ISBN 9781402054082.
\newblock \doi{10.1107/97809553602060000600}.
\newblock URL \url{https://it.iucr.org/Cb/ch6o1v0001/}.

\bibitem[Grin et~al.(1991)Grin, Rogl, Chevalier, Fedorchuk, and Gryniv]{CeGaGe_Grin}
Yu.N. Grin, P.~Rogl, B.~Chevalier, A.A. Fedorchuk, and I.A. Gryniv.
\newblock Physical properties of binary cerium gallides and ternary cerium-germanium gallides.
\newblock \emph{Journal of the Less Common Metals}, 167\penalty0 (2):\penalty0 365--371, 1991.
\newblock \doi{10.1016/0022-5088(91)90289-G}.

\bibitem[P\"{o}ttgen and Chevalier(2015)]{CeXX_GaGe_Pott}
Rainer P\"{o}ttgen and Bernard Chevalier.
\newblock Equiatomic cerium intermetallics {Ce$XX'$} with two p elements.
\newblock \emph{Zeitschrift f\"{u}r Naturforschung B}, 70\penalty0 (10):\penalty0 695--704, 2015.
\newblock \doi{doi:10.1515/znb-2015-0109}.
\newblock URL \url{https://doi.org/10.1515/znb-2015-0109}.

\bibitem[Dhar et~al.(1993)Dhar, Pattalwar, and Vijayaraghavan]{CeGaGe_Dhar}
S.K. Dhar, S.M. Pattalwar, and R.~Vijayaraghavan.
\newblock {CeGeGa}—a ferromagnetic dense {K}ondo system.
\newblock \emph{Physica B: Condensed Matter}, 186–188:\penalty0 491–493, may 1993.
\newblock ISSN 0921-4526.
\newblock \doi{10.1016/0921-4526(93)90613-b}.
\newblock URL \url{http://dx.doi.org/10.1016/0921-4526(93)90613-B}.

\bibitem[Ram et~al.(2023)Ram, Malick, Hossain, and Kaczorowski]{CeGaGe_andPrGaGe_Ram_RPB_2023}
Daloo Ram, Sudip Malick, Zakir Hossain, and Dariusz Kaczorowski.
\newblock Magnetic, thermodynamic, and magnetotransport properties of {CeGaGe} and {PrGaGe} single crystals.
\newblock \emph{Phys. Rev. B}, 108:\penalty0 024428, Jul 2023.
\newblock \doi{10.1103/PhysRevB.108.024428}.
\newblock URL \url{https://link.aps.org/doi/10.1103/PhysRevB.108.024428}.

\bibitem[Rietveld(1969)]{RietveldRefinement}
H.~M. Rietveld.
\newblock A profile refinement method for nuclear and magnetic structures.
\newblock \emph{Journal of Applied Crystallography}, 2\penalty0 (2):\penalty0 65–71, June 1969.
\newblock ISSN 0021-8898.
\newblock \doi{10.1107/s0021889869006558}.
\newblock URL \url{http://dx.doi.org/10.1107/S0021889869006558}.

\bibitem[Sears(1992)]{NeutronScatteringLengths}
Varley~F. Sears.
\newblock Neutron scattering lengths and cross sections.
\newblock \emph{Neutron News}, 3\penalty0 (3):\penalty0 26--37, 1992.
\newblock \doi{10.1080/10448639208218770}.
\newblock URL \url{https://doi.org/10.1080/10448639208218770}.

\bibitem[Neu(2021)]{NeutronScatteringLengths_URL}
{N}eutron scattering lengths and cross sections --- ncnr.nist.gov.
\newblock \url{https://www.ncnr.nist.gov/resources/n-lengths/}, 2021.
\newblock [Accessed 16-07-2024].

\bibitem[Sup()]{SupplementalMaterialForThisManuscript}
\emph{See Supplemental Material at [link.aps.org/supplemental/10.1103/PhysRevB.111.184102] for details of the energy dispersive X-ray spectroscopy; details of the powder X-ray diffraction; details of the single crystal X-ray diffraction; further verification of single crystal neutron diffraction data quality; and the quality of the data where \mbox{$10^{-4}\leq |F_c|^2 \leq 1.25$} for the $I4_1/amd$ model}.

\bibitem[Statham(2007)]{PulsePileUpCorrection}
P~Statham.
\newblock {Digital Pulse Processing and Pile Up Correction for Accurate Interpretation of High Rate {SDD} Spectrum Images}.
\newblock \emph{Microscopy and Microanalysis}, 13\penalty0 (S02):\penalty0 1428--1429, 08 2007.
\newblock ISSN 1431-9276.
\newblock \doi{10.1017/S1431927607073242}.
\newblock URL \url{https://doi.org/10.1017/S1431927607073242}.

\bibitem[Rodríguez-Carvajal(1993)]{FullProf}
Juan Rodríguez-Carvajal.
\newblock Recent advances in magnetic structure determination by neutron powder diffraction.
\newblock \emph{Physica B: Condensed Matter}, 192\penalty0 (1–2):\penalty0 55–69, October 1993.
\newblock ISSN 0921-4526.
\newblock \doi{10.1016/0921-4526(93)90108-i}.
\newblock URL \url{http://dx.doi.org/10.1016/0921-4526(93)90108-I}.

\bibitem[Aroyo et~al.(2016)Aroyo, Burzlaff, Chapuis, Fischer, Flack, Glazer, Grimmer, Gruber, Hahn, Klapper, Koch, Kostantinov, Kopsk\'y, Litvin, Looijenga-Vos, Momma, M\"uller, Shmueli, Souvignier, Spence, de~Wolff, Wondratschek, and Zimmermann]{IUCR_SpaceGroups}
M.~I. Aroyo, H.~Burzlaff, G.~Chapuis, W.~Fischer, H.~D. Flack, A.~M. Glazer, H.~Grimmer, B.~Gruber, Th. Hahn, H.~Klapper, E.~Koch, P.~Kostantinov, V.~Kopsk\'y, D.~B. Litvin, A.~Looijenga-Vos, K.~Momma, U.~M\"uller, U.~Shmueli, B.~Souvignier, J.~C.~H. Spence, P.~M. de~Wolff, H.~Wondratschek, and H.~Zimmermann.
\newblock \emph{International Tables for Crystallography Volume {A}: Space-group symmetry}, chapter 2.3.
\newblock International Union of Crystallography, 2016.

\bibitem[Coates et~al.(2018)Coates, Cao, Chakoumakos, Frontzek, Hoffmann, Kovalevsky, Liu, Meilleur, dos Santos, Myles, Wang, and Ye]{ORNL_instruments2018}
L.~Coates, H.~B. Cao, B.~C. Chakoumakos, M.~D. Frontzek, C.~Hoffmann, A.~Y. Kovalevsky, Y.~Liu, F.~Meilleur, A.~M. dos Santos, D.~A.~A. Myles, X.~P. Wang, and F.~Ye.
\newblock A suite-level review of the neutron single-crystal diffraction instruments at {O}ak {R}idge {N}ational {L}aboratory.
\newblock \emph{Review of Scientific Instruments}, 89\penalty0 (9), September 2018.
\newblock ISSN 1089-7623.
\newblock \doi{10.1063/1.5030896}.
\newblock URL \url{http://dx.doi.org/10.1063/1.5030896}.

\bibitem[Akeroyd et~al.(2013)Akeroyd, Ansell, Antony, Arnold, Bekasovs, Bilheux, Borreguero, Brown, Buts, Campbell, Champion, Chapon, Clarke, Cottrell, Dalgliesh, Dillow, Doucet, Draper, Fowler, Gigg, Granroth, Hagen, Heller, Hillier, Howells, Jackson, Kachere, Koennecke, Le~Bourlot, Leal, Lynch, Manuel, Markvardsen, McGreevy, Mikkelson, Mikkelson, Miller, Nagella, Nielsen, Palmen, Parker, Pascal, Passos, Perring, Peterson, Pratt, Proffen, Radaelli, Rainey, Ren, Reuter, Sastry, Savici, Taylor, Taylor, Thomas, Tolchenov, Whitley, Whitty, Williams, Zhou, and Zikovsky]{Mantid}
Freddie Akeroyd, Stuart Ansell, Sofia Antony, Owen Arnold, Arturs Bekasovs, Jean Bilheux, Jose Borreguero, Keith Brown, Alex Buts, Stuart Campbell, Dickon Champion, Laurent Chapon, Matt Clarke, Stephen Cottrell, Robert Dalgliesh, David Dillow, Mathieu Doucet, Nick Draper, Ronald Fowler, Martyn~A. Gigg, Garrett Granroth, Mark Hagen, William Heller, Adrian Hillier, Spencer Howells, Samuel Jackson, Dereck Kachere, Mark Koennecke, Christophe Le~Bourlot, Ricardo Leal, Vickie Lynch, Pascal Manuel, Anders Markvardsen, Robert McGreevy, Dennis Mikkelson, Ruth Mikkelson, Ross Miller, Sri Nagella, Torben Nielsen, Karl Palmen, Peter~G. Parker, Manuel Pascal, Gesner Passos, Toby Perring, Peter~F. Peterson, Francis Pratt, Thomas Proffen, Paolo Radaelli, Jay Rainey, Shelly Ren, Michael Reuter, Lakshmi Sastry, Andrei Savici, Jon Taylor, Russell~J. Taylor, Mike Thomas, Roman Tolchenov, Robert Whitley, Michael Whitty, Steve Williams, Wenduo Zhou, and Janik Zikovsky.
\newblock Mantid: Manipulation and analysis toolkit for instrument data., 2013.
\newblock URL \url{http://www.mantidproject.org}.

\bibitem[Applin et~al.(2024)Applin, Baust, Diaz-Alvarez, Finn, Foxley, Haigh, Hampson, and Peterson]{MantidRelease}
Robert Applin, Rachel Baust, Adrian Diaz-Alvarez, Caila Finn, Sarah Foxley, Jonathan Haigh, Thomas Hampson, and Peter~F. Peterson.
\newblock Mantid 6.9.1: Manipulation and analysis toolkit for instrument data., 2024.
\newblock URL \url{http://docs.mantidproject.org/v6.9.1/release/v6.9.1/index.html}.

\bibitem[Toby and Von~Dreele(2013)]{GSAS}
Brian~H. Toby and Robert~B. Von~Dreele.
\newblock {GSAS-II}: the genesis of a modern open-source all purpose crystallography software package.
\newblock \emph{Journal of Applied Crystallography}, 46\penalty0 (2):\penalty0 544–549, March 2013.
\newblock ISSN 0021-8898.
\newblock \doi{10.1107/s0021889813003531}.
\newblock URL \url{http://dx.doi.org/10.1107/S0021889813003531}.

\bibitem[Mughabghab(2018)]{AtlasOfNeutronResonances}
S~F Mughabghab.
\newblock Thermal cross sections.
\newblock In \emph{Atlas of Neutron Resonances}, pages 1--16. Elsevier, 2018.

\bibitem[nis()]{nistThermalNeutron}
{T}hermal neutron resonances --- ncnr.nist.gov.
\newblock \url{https://www.ncnr.nist.gov/resources/activation/resonance.html}.
\newblock [Accessed 14-11-2024].

\bibitem[{Bruker AXS Inc.}(2016)]{BrukerApex}
{Bruker AXS Inc.}
\newblock {APEX3}, {SAINT}, and {SADABS}.
\newblock \emph{Madison, WI, USA}, 2016.

\bibitem[Krause et~al.(2015)Krause, Herbst-Irmer, and Stalke]{SADABS}
Lennard Krause, Regine Herbst-Irmer, and Dietmar Stalke.
\newblock An empirical correction for the influence of low-energy contamination.
\newblock \emph{Journal of Applied Crystallography}, 48\penalty0 (6):\penalty0 1907–1913, November 2015.
\newblock ISSN 1600-5767.
\newblock \doi{10.1107/s1600576715020440}.
\newblock URL \url{http://dx.doi.org/10.1107/S1600576715020440}.

\bibitem[Sheldrick(2015{\natexlab{a}})]{SHELXT_A}
George~M. Sheldrick.
\newblock {SHELXT}– integrated space-group and crystal-structure determination.
\newblock \emph{Acta Crystallographica Section A Foundations and Advances}, 71\penalty0 (1):\penalty0 3–8, January 2015{\natexlab{a}}.
\newblock ISSN 2053-2733.
\newblock \doi{10.1107/s2053273314026370}.
\newblock URL \url{http://dx.doi.org/10.1107/S2053273314026370}.

\bibitem[Sheldrick(2015{\natexlab{b}})]{SHELXL_C}
George~M. Sheldrick.
\newblock Crystal structure refinement with {SHELXL}.
\newblock \emph{Acta Crystallographica Section C Structural Chemistry}, 71\penalty0 (1):\penalty0 3–8, January 2015{\natexlab{b}}.
\newblock ISSN 2053-2296.
\newblock \doi{10.1107/s2053229614024218}.
\newblock URL \url{http://dx.doi.org/10.1107/S2053229614024218}.

\bibitem[Flack(1983)]{Flack}
H.~D. Flack.
\newblock {On enantiomorph-polarity estimation}.
\newblock \emph{Acta Crystallographica Section A}, 39\penalty0 (6):\penalty0 876--881, Nov 1983.
\newblock \doi{10.1107/S0108767383001762}.
\newblock URL \url{https://doi.org/10.1107/S0108767383001762}.

\bibitem[W\"{o}hler and Deville(1857)]{WohlerFlux_1857}
F.~W\"{o}hler and H.~Sainte‐Claire Deville.
\newblock Ueber das bor.
\newblock \emph{Justus Liebigs Annalen der Chemie}, 101\penalty0 (3):\penalty0 347–355, January 1857.
\newblock ISSN 0075-4617.
\newblock \doi{10.1002/jlac.18571010317}.
\newblock URL \url{http://dx.doi.org/10.1002/jlac.18571010317}.

\bibitem[Canfield and Fisk(1992)]{CanfieldFlux}
P.~C. Canfield and Z.~Fisk.
\newblock Growth of single crystals from metallic fluxes.
\newblock \emph{Philosophical Magazine B}, 65\penalty0 (6):\penalty0 1117–1123, June 1992.
\newblock ISSN 1463-6417.
\newblock \doi{10.1080/13642819208215073}.
\newblock URL \url{http://dx.doi.org/10.1080/13642819208215073}.

\bibitem[Valent\'in‐P\'erez et~al.(2021)Valent\'in‐P\'erez, Rosa, Hillard, and Giorgi]{ChiralityDetInCryst_Perez2021}
\'Angela Valent\'in‐P\'erez, Patrick Rosa, Elizabeth~A. Hillard, and Michel Giorgi.
\newblock Chirality determination in crystals.
\newblock \emph{Chirality}, 34\penalty0 (2):\penalty0 163--181, November 2021.
\newblock ISSN 1520-636X.
\newblock \doi{10.1002/chir.23377}.
\newblock URL \url{http://dx.doi.org/10.1002/chir.23377}.

\bibitem[Parsons(2017)]{DetAbsConfig_Parsons_2017}
Simon Parsons.
\newblock Determination of absolute configuration using {X}-ray diffraction.
\newblock \emph{Tetrahedron: Asymmetry}, 28\penalty0 (10):\penalty0 1304–1313, October 2017.
\newblock ISSN 0957-4166.
\newblock \doi{10.1016/j.tetasy.2017.08.018}.
\newblock URL \url{http://dx.doi.org/10.1016/j.tetasy.2017.08.018}.

\bibitem[Flandorfer et~al.(1998)Flandorfer, Kaczorowski, Gr\"{o}bner, Rogl, Wouters, Godart, and Kostikas]{CeAl_SiGe_Flandorfer_JoSSC_1998}
H.~Flandorfer, D.~Kaczorowski, J.~Gr\"{o}bner, P.~Rogl, R.~Wouters, C.~Godart, and A.~Kostikas.
\newblock The systems {Ce}–{Al}–({Si}, {Ge}): Phase equilibria and physical properties.
\newblock \emph{Journal of Solid State Chemistry}, 137\penalty0 (2):\penalty0 191–205, May 1998.
\newblock ISSN 0022-4596.
\newblock \doi{10.1006/jssc.1997.7660}.
\newblock URL \url{http://dx.doi.org/10.1006/jssc.1997.7660}.

\end{thebibliography}


\begin{thebibliography}{1}
\providecommand{\natexlab}[1]{#1}
\providecommand{\url}[1]{\texttt{#1}}
\expandafter\ifx\csname urlstyle\endcsname\relax
  \providecommand{\doi}[1]{doi: #1}\else
  \providecommand{\doi}{doi: \begingroup \urlstyle{rm}\Url}\fi

\bibitem[Statham(2007)]{PulsePileUpCorrection}
P~Statham.
\newblock {Digital Pulse Processing and Pile Up Correction for Accurate Interpretation of High Rate SDD Spectrum Images}.
\newblock \emph{Microscopy and Microanalysis}, 13\penalty0 (S02):\penalty0 1428--1429, 08 2007.
\newblock ISSN 1431-9276.
\newblock \doi{10.1017/S1431927607073242}.
\newblock URL \url{https://doi.org/10.1017/S1431927607073242}.

\end{thebibliography}
\end{document}


\title{Structural characterization of the candidate Weyl semimetal \mbox{CeGaGe} - Supplemental Material}
\author{Liam J. Scanlon}
\author{Santosh Bhusal}
\affiliation{Department of Physics and Astronomy, University of Kentucky, Lexington, KY 40506 USA}
\author{Christina M. Hoffmann}
\author{Junhong He}
\affiliation{Neutron Scattering Division, Oak Ridge National Laboratory, Oak Ridge, TN 37831 USA}
\author{Sean R. Parkin}
\affiliation{Department of Chemistry, University of Kentucky, Lexington, KY 40506 USA}
\author{Brennan J. Arnold}
\author{William J. Gannon}
\affiliation{Department of Physics and Astronomy, University of Kentucky, Lexington, KY 40506 USA}
\date{\today}
\maketitle
\FloatBarrier
\section{Energy Dispersive X-Ray Spectroscopy}
    
    \FloatBarrier

\begin{figure}[h]
            \includegraphics[width=\linewidth]{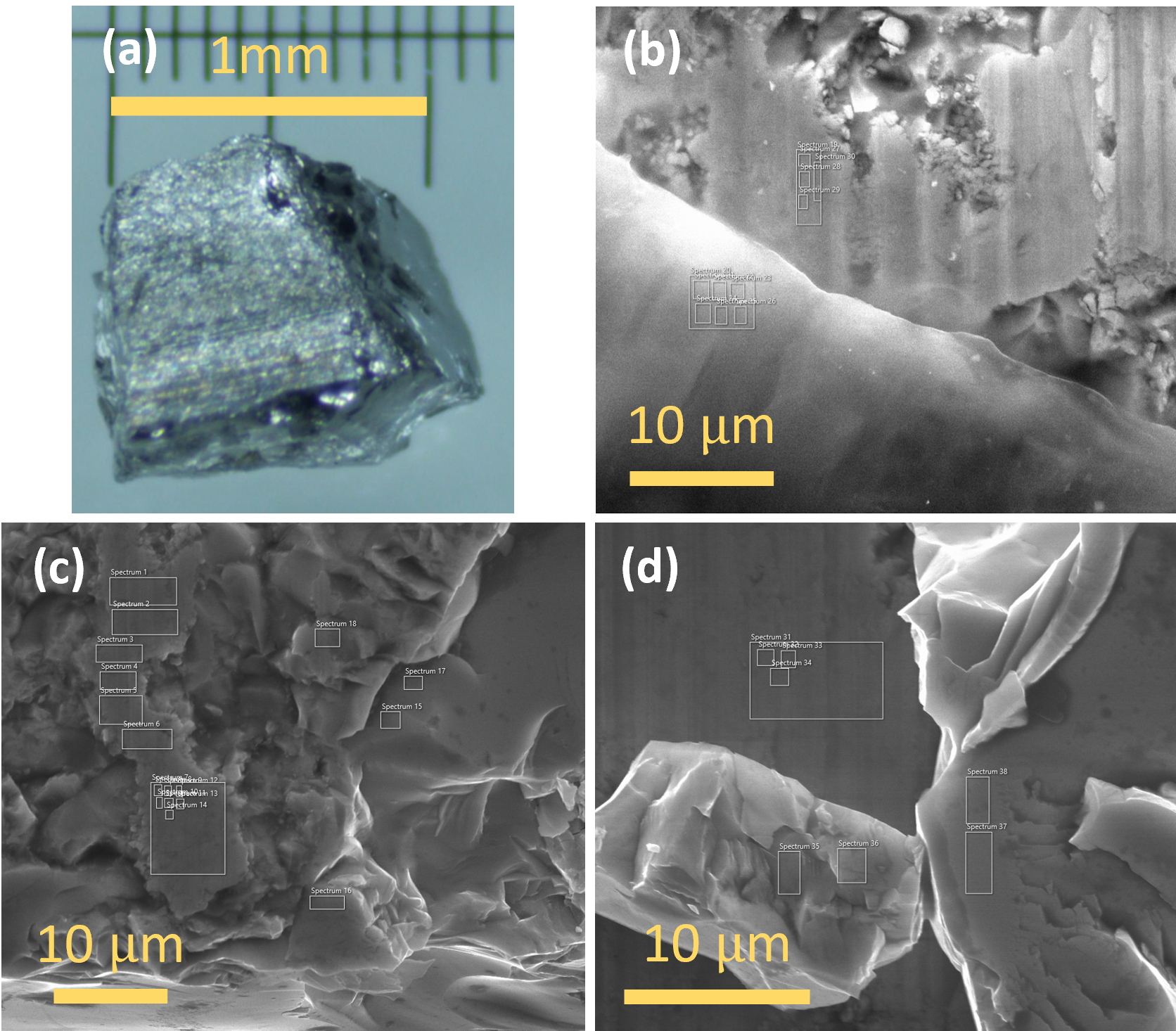}
		\caption{Images of the CeGaGe sample used for SNCD:  (a) Optical image of the sample.  (b-d) Scanning electron images of the sample from the Quanta FEI electron microscope. In each image, the faint white boxes indicate the sites where EDX data were taken.
		}
		\label{fig:images}
	\end{figure}

\begin{figure*}
        \begin{centering}
            \includegraphics[width=\linewidth]{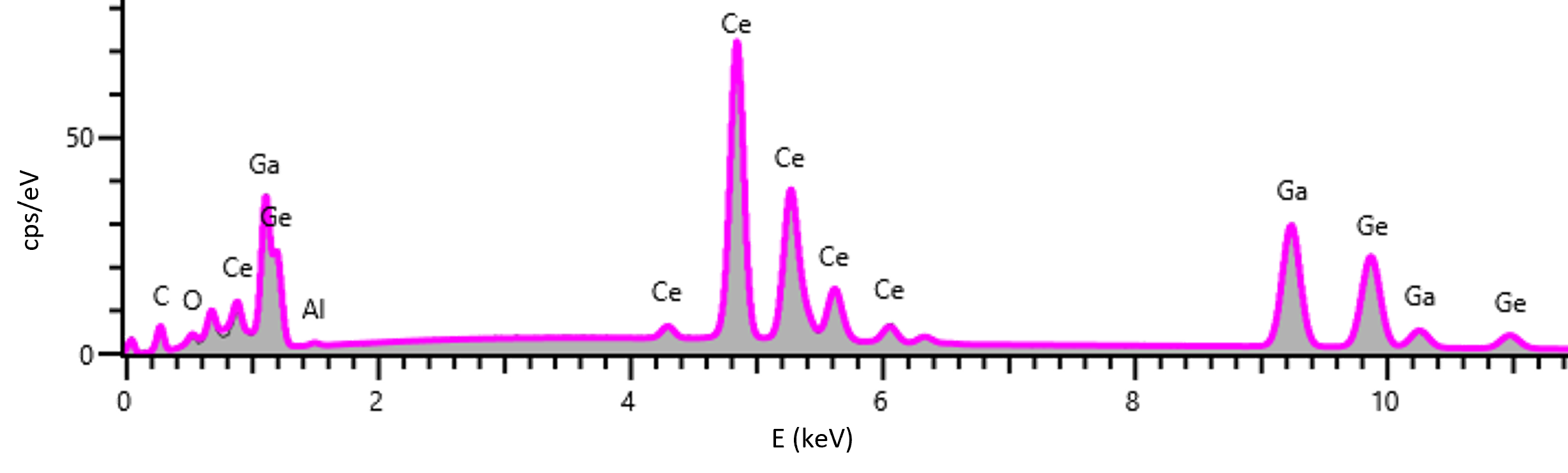}
        \end{centering}
        \caption{
        Representative EDX spectrum for CeGaGe. The gray, solid region represents the data and the pink curve represents a fit to the data with pulse pileup correction \cite{PulsePileUpCorrection}.  The elements responsible for the main transitions observed are labeled. 
        }
        \label{fig:representativeEDXSpectrum}
    \end{figure*}

An optical image and several scanning electron images of the sample used for single crystal neutron diffraction (SNCD) can be seen in Fig. \ref{fig:images}.  The optical image (Fig. \ref{fig:images}(a)) shows the relative scale of the sample used in the neutron diffraction experiment. The scanning electron images of three different areas of the sample (Fig. \ref{fig:images}(b-d)) show the sample surface in more detail.  White boxes in the scanning electron images mark all 38 sites where energy dispersive x-ray spectroscopy (EDX) data were taken. A representative EDX spectrum is shown in Fig. \ref{fig:representativeEDXSpectrum}. The fit to the EDX spectrum includes the pulse pileup correction, so photon counts that overlap in time can be resolved individually \cite{PulsePileUpCorrection}.  The fit matches the data well.  When oxygen is included in the stoichiometric calculation, the average oxygen content is 15.67\% with a standard deviation of 11.51\% for all 38 sites.  As there is no evidence of any oxide in any of the diffraction experiments conducted on these samples, this indicates that the oxygen present in the EDX spectra is residual oxygen in the SEM chamber.\commented{We also note a small amount of carbon present from the tape that is typically used to mount the sample in the scanning electron microscope.} Uncertainty values for EDX were were found by taking the standard deviation of the compositions found at 38 sites on the sample.

 \vspace{0.2cm}
	\section{Powder X-Ray Diffraction}
 
    The refined crystallographic parameters for both the $I4_1md$ and $I4_1/amd$  structural models from powder x-ray diffraction (PXRD) measurements can be seen in table \ref{tab:cifStatsPXRD}.  The constraint that the fractional $z$-coordinates of Ga and Ge add to 1 in the $I4_1md$ structure was lifted, resulting in the fractional $z$-coordinates summing to $0.996(2)$.  When a SCND refinement with $I4_1md$ synmetry was carried out without this constraint, the $z$-coordinates added to $0.999(1)$. From the  SCND refinement, it was concluded that the $z$-coordinates very likely add to 1 and the discrepancy from this in the refinement was an artifact of numerical minimization and so a refinement with the $z$-coordinate constraint is reported in the main text.  Given table \ref{tab:cifStatsPXRD}, the reliability factors and goodness of fit values reported in the main text, and Fig. \ref{fig:pxrd} in the main text, it can bee seen that the PXRD refinements do not distinguish between the two space groups.\commented{ considering the similarity of the fits, fit statistics, and refined parameters.}\\
    
    \begin{table}
    \setlength\extrarowheight{2pt}
    \sisetup{group-digits=false}
		\begin{center}
			\begin{tblr}{
					colspec = {|c|c|c|c|}
				}
				\hline
				& $a$ $\left[\text{\r{A}}\right]$ & $c$ $\left[\text{\r{A}}\right]$ & $V$ $\left[\text{\r{A}}^3\right]$  \\ \hline
				$I4_1md$  & 4.27503(2) & 14.5700(1) & 266.280(3)  \\ \hline
				$I4_1/amd$ & 4.27505(2) & 14.5700(1) & 266.283(3)  \\ \hline
			\end{tblr}
			~\\
			\vspace{0.1cm}
			~\\
			\begin{tabular}{
					|c|c|
                    S[table-format = 1.1]|
                    S[table-format = 1.1]|
                    S[table-format = 1.4(1)]
                    |S[table-format = 1.2]|
				}
				\hline
				 \ & Atom & $x$ & $y$ & $z$ & {occ.} \\ \hline
				$I4_1md$  & Ce & 0.0 & 0.0 & 0.000 & 1.00 \\ \hline
				$I4_1/amd$ & Ce & 0.0 & 0.0 & 0.000 & 1.00 \\ \hline
				$I4_1md$  & Ga & 0.0 & 0.0 & 0.4155(9) & 1.00 \\ \hline
				$I4_1/amd$ & Ga & 0.0 & 0.0 & 0.4177(1) & 0.50 \\ \hline
				$I4_1md$  & Ge & 0.0 & 0.0 & 0.5803(8) & 1.00 \\ \hline
				$I4_1/amd$ & Ge & 0.0 & 0.0 & 0.4177(1) & 0.50 \\ \hline
			\end{tabular}
		\end{center}
		\caption{
			Crystallographic information of the \mbox{CeGaGe} structure using both  $I4_1md$ and $I4_1/amd$ symmetries in PXRD refinements.  The crystallographic axes of the two space group refinements are within uncertainty of each other and atomic coordinates match to two significant figures.
		}
		\label{tab:cifStatsPXRD}
	\end{table}

 \vspace{0.2cm}
    \section{Single Crystal X-Ray Diffraction}
    
  Crystallographic parameters for the structure of the zone refined sample used for SCND, determined by single crystal x-ray diffraction (SCXRD) at 100 K, are summarized in Table \ref{tab:cifStatsSCXRD}.  SCXRD measurements were made on a crystal grown with the flux technique at both room temperature and at $T=100$~K.  There is a subtle, but clear structural transition that is observed in the flux-grown sample, as described in the main text.  A refinement of the 100 K SCXRD data for this sample was carried out  using $P4_3$ symmetry (space group 78).  The crystallographic information for this refinement is in table \ref{tab:SCXRD_P43}.

    \begin{table}
    \setlength\extrarowheight{2pt}
    \sisetup{group-digits=false}
		\begin{center}
			\begin{tblr}{
					colspec = {|c|c|c|c|}
				}
				\hline
				& $a$ $\left[\text{\r{A}}\right]$ & $c$ $\left[\text{\r{A}}\right]$ & $V$ $\left[\text{\r{A}}^3\right]$  \\ \hline
				$I4_1md$  & 4.2673(1) & 14.5413(4) & 264.80(1)  \\ \hline
				$I4_1/amd$ & 4.2752(1) & 14.4999(6) & 265.02(2)  \\ \hline
			\end{tblr}
			~\\
			\vspace{0.1cm}
			~\\
			\begin{tabular}{
					|c|c|
                    S[table-format = 1.1]|
                    S[table-format = 1.1]|
                    S[table-format = 1.4(1)]
                    |S[table-format = 1.2]|
				}
				\hline
				& Atom & $x$ & $y$ & $z$ & {occ.} \\ \hline
				$I4_1md$  & Ce & 0.0 & 0.0 & 0.0 & 1.00 \\ \hline
				$I4_1/amd$ & Ce & 0.0 & 0.0 & 0.0000 & 1.00 \\ \hline
				$I4_1md$  & Ga & 0.0 & 0.0 & 0.4159(1) & 1.00 \\ \hline
				$I4_1/amd$ & Ga & 0.0 & 0.0 & 0.417(2) & 0.50 \\ \hline
				$I4_1md$  & Ge & 0.0 & 0.0 & 0.5825(1) & 1.00 \\ \hline
				$I4_1/amd$ & Ge & 0.0 & 0.0 & 0.417(2) & 0.50 \\ \hline
			\end{tabular}
		\end{center}
		\caption{
			Crystallographic information of the \mbox{CeGaGe} structure using both $I4_1md$ and $I4_1/amd$ symmetries in SCXRD refinements.  The crystallographic axes of the two space group refinements match to two significant figures and the atomic coordinates match to two significant figures.
		}
		\label{tab:cifStatsSCXRD}
	\end{table}

\begin{table}[]
    \setlength\extrarowheight{2pt}
    \sisetup{group-digits=false}
    \begin{center}
    \begin{tabular}{|c|c|c|}
    \hline
    $a$ $\left[\text{\r{A}}\right]$ & $c$ $\left[\text{\r{A}}\right]$ & $V$ $\left[\text{\r{A}}^3\right]$  \\ \hline
    4.2756(1) & 14.5414(3) & 265.83(1) \\ \hline
    \end{tabular}
    ~\\
    \vspace{0.1cm}
    ~\\
    \resizebox{\columnwidth}{!}{
    \begin{tabular}{
					|c|
                    S[table-format = 1.5(1)]|
                    S[table-format = 1.5(1)]|
                    S[table-format = 1.5(1)]|
                    S[table-format = 1.2]|
				}
         \hline
         Atom & {$x$} & {$y$} & {$z$} & {occ.} \\ \hline
         Ce & 0.24849(8) & 0.25156(8) & 0.66595(9) & 1.00 \\ \hline
         Ga & 0.2530(2) & 0.7297(3) & 0.50001(4) & 1.00 \\ \hline
         Ge & 0.2297(3) & 0.7531(2) & 0.33335(4) & 1.00 \\ \hline
    \end{tabular}
    }
    \end{center}
    
    \caption{Crystallographic information of the CeGaGe structure for $P4_3$ single crystal XRD refinements.}
    \label{tab:SCXRD_P43}
\end{table}

    \vspace{0.2cm}
	\section{Verifying SCND data quality}
	
  \begin{figure*}
		\begin{center}
			\includegraphics[angle=90, width=0.49\linewidth]{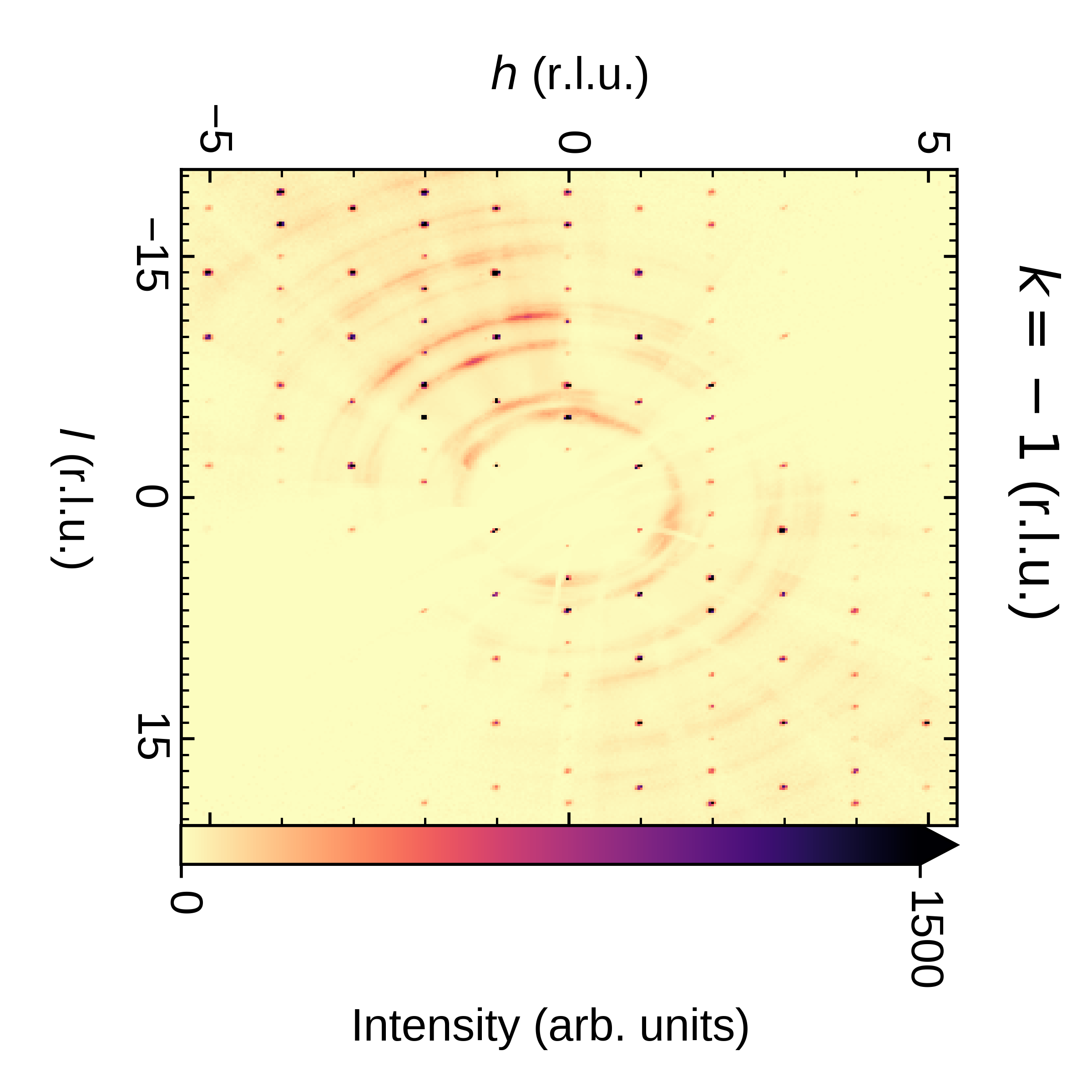}
			\includegraphics[angle=90, width=0.49\linewidth]{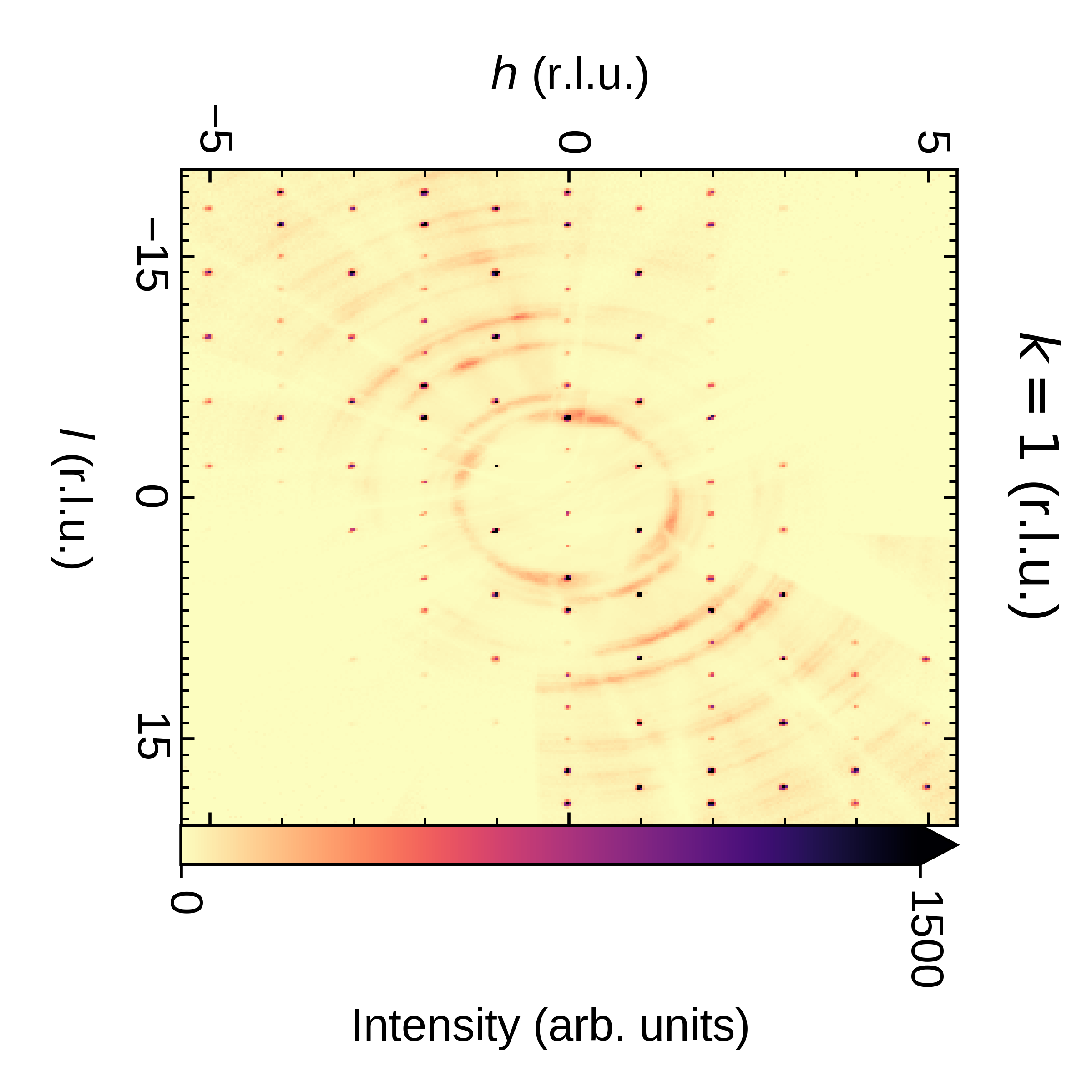}
		\end{center}
		\caption{Plots of merged SCND data in the $k=\pm1$ plane.  Data for the 19 different goniometer angles merged well and sample is single crystalline.}
		\label{fig:KeqPm1DataSlices}
\end{figure*}	
 \begin{figure*}
    \begin{center}
        \begin{minipage}{0.4\linewidth}
            \includegraphics[height=1.1in, right]{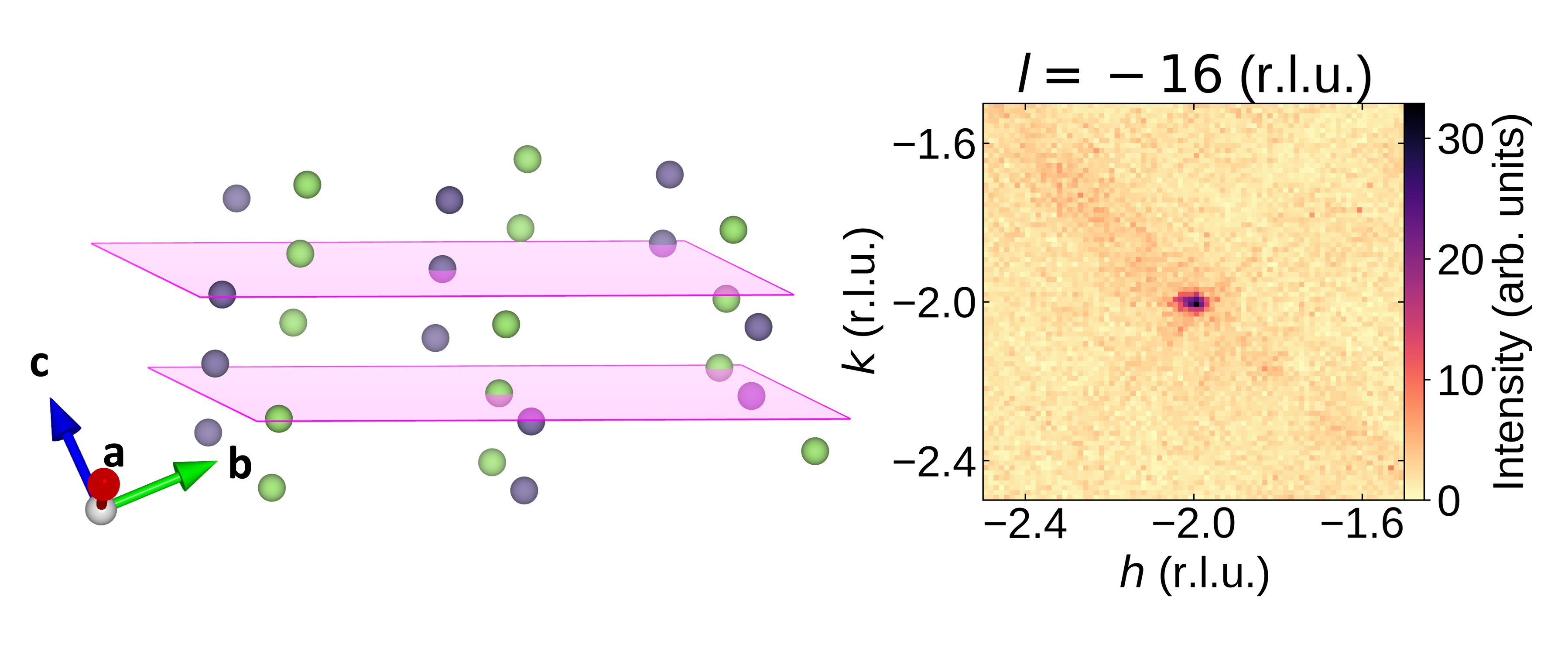}
	\end{minipage}
        \hspace{0.1\linewidth}
        \begin{minipage}{0.4\linewidth}
            \includegraphics[height=1.1in, right]{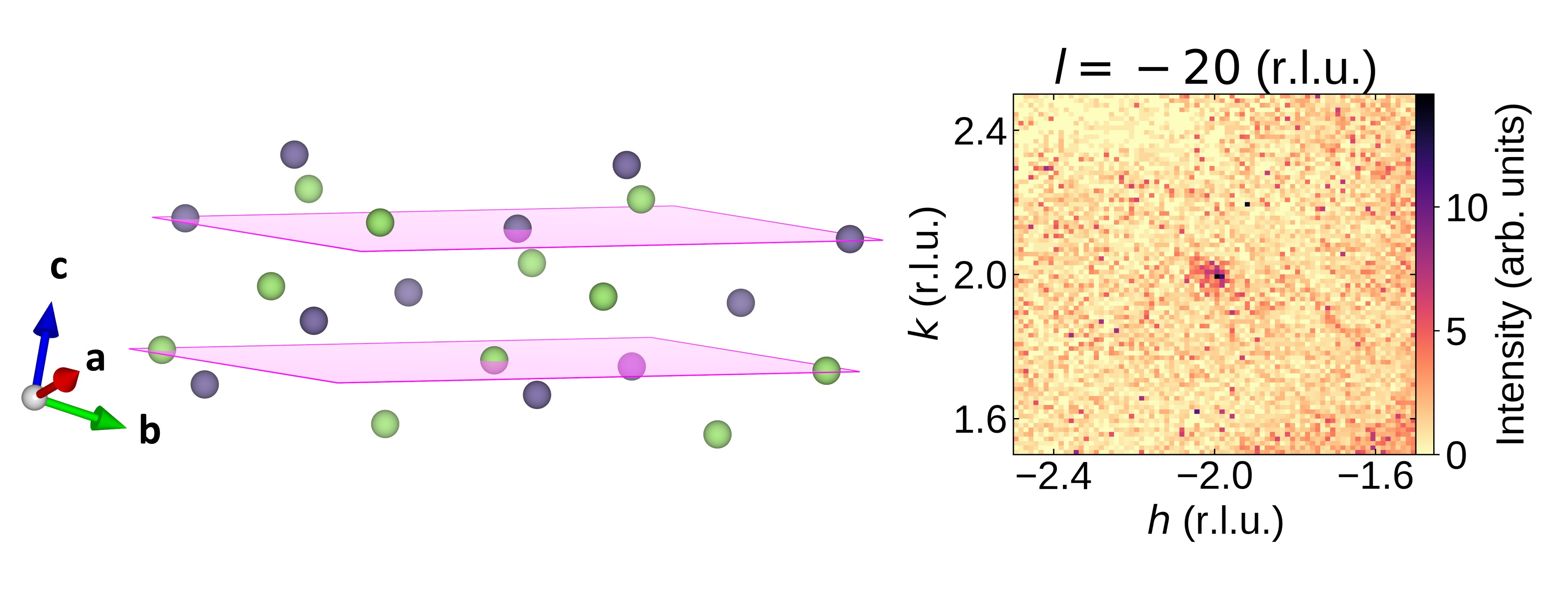}
        \end{minipage}
        \begin{minipage}{0.4\linewidth}
		    \includegraphics[height=1.1in, right]{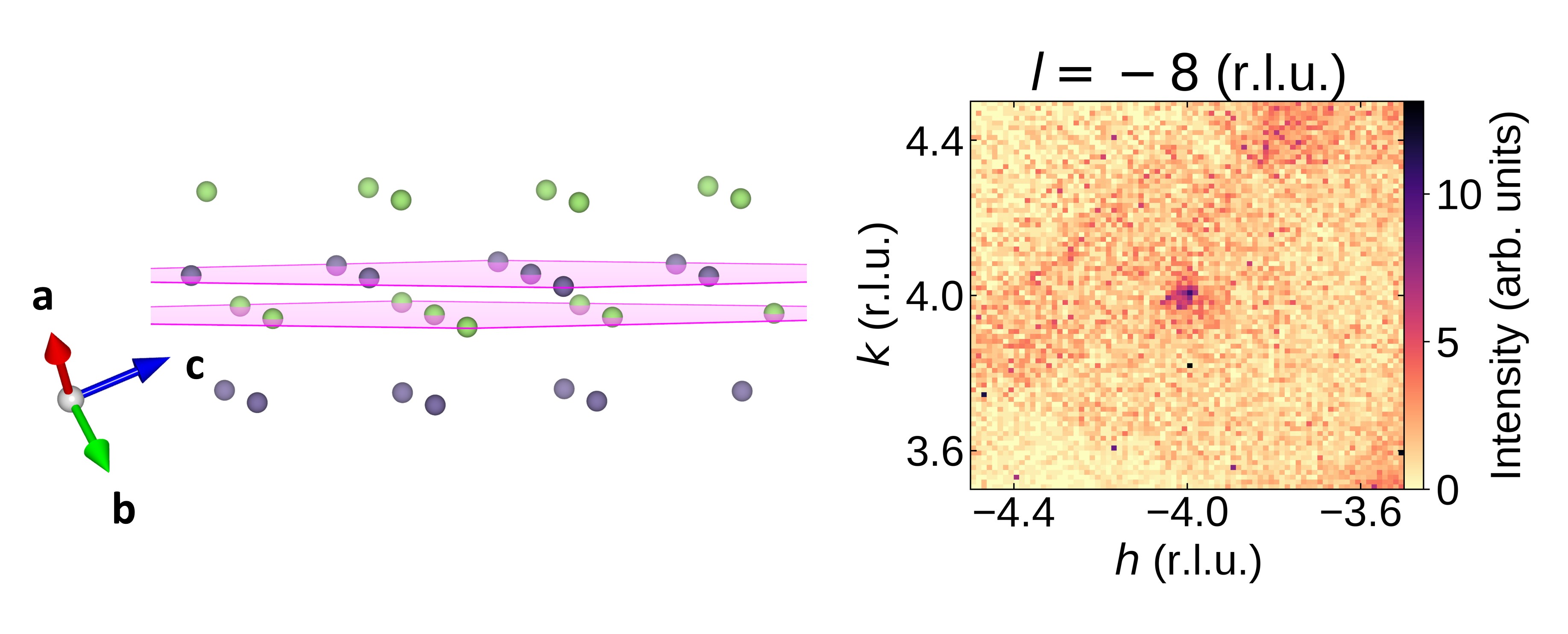}
        \end{minipage}
        \hspace{0.1\linewidth}
        \begin{minipage}{0.4\linewidth}
		    \includegraphics[height=1.1in, right]{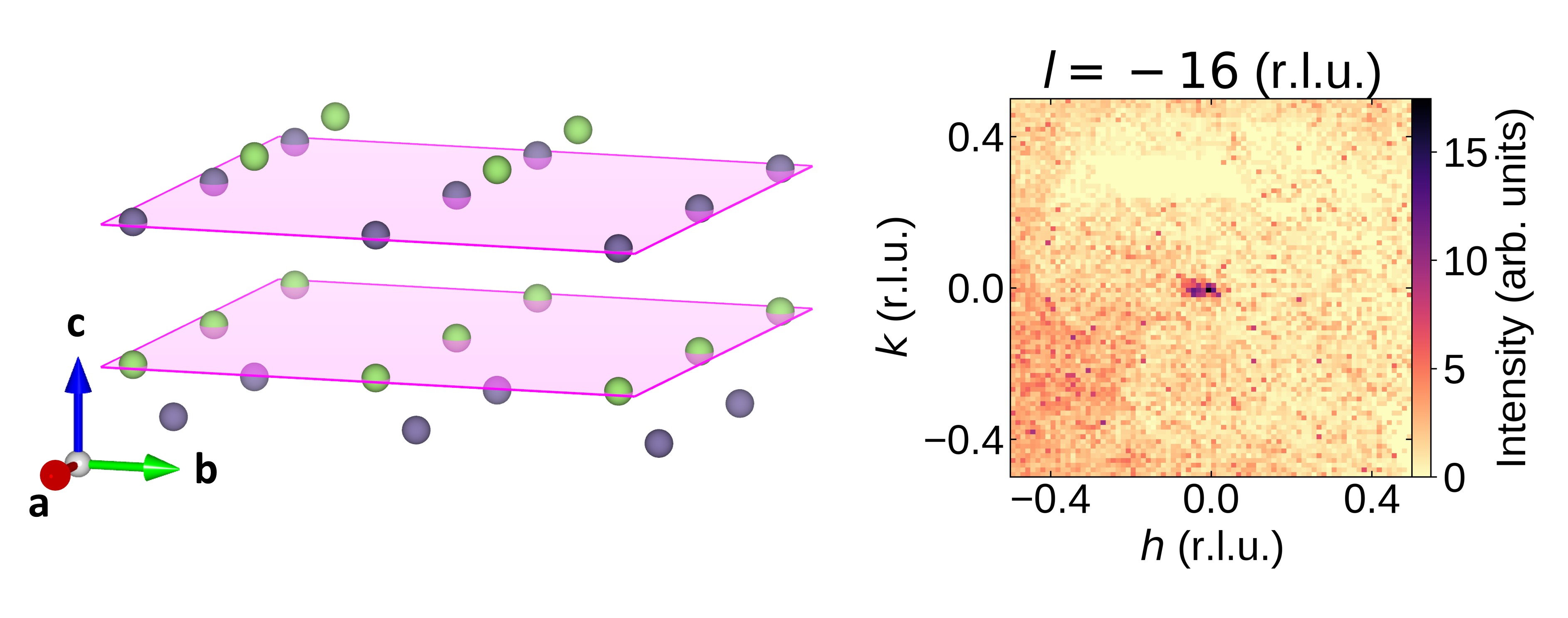}
        \end{minipage}
        \hspace{0.1\linewidth}
	  \begin{minipage}{0.4\linewidth}	
            \includegraphics[height=1.1in, right]{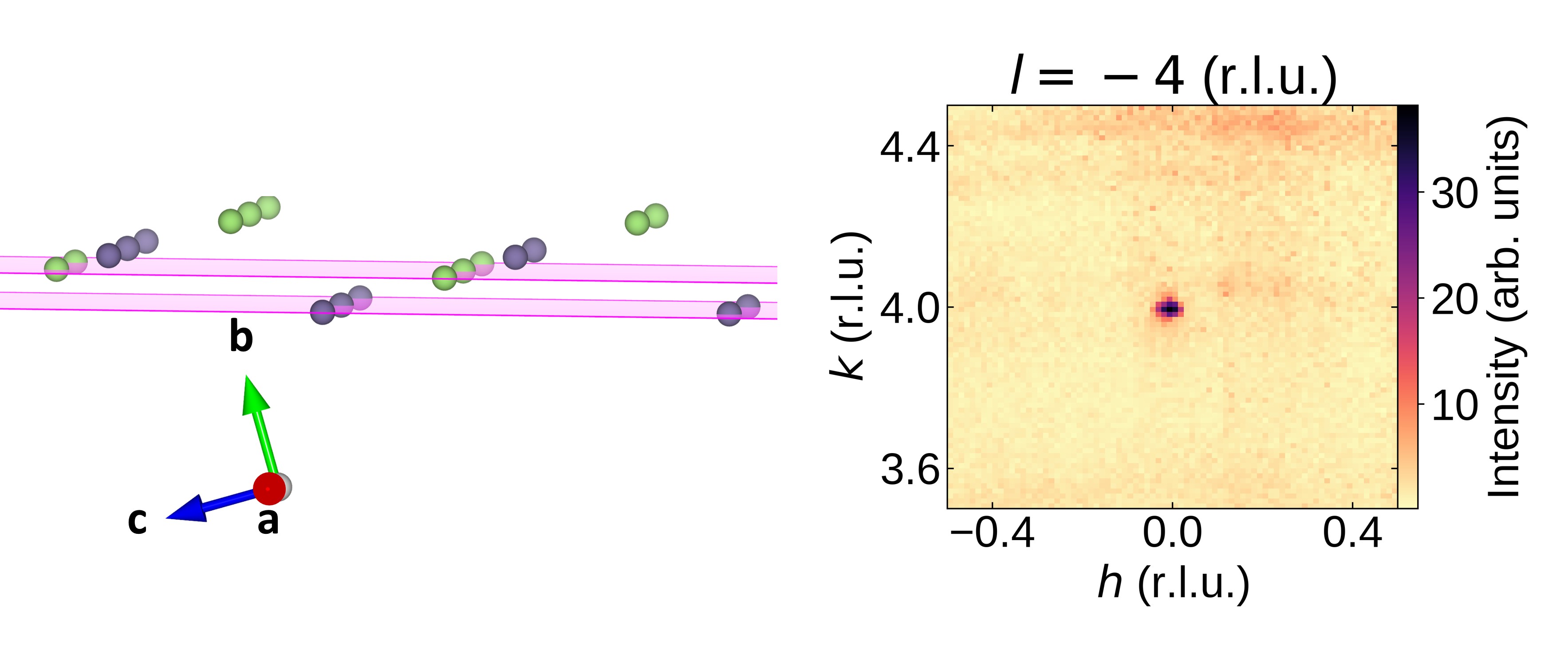}
        \end{minipage}
        \hspace{0.1\linewidth}
	    \begin{minipage}{0.4\linewidth}	
            \includegraphics[height=1.1in, right]{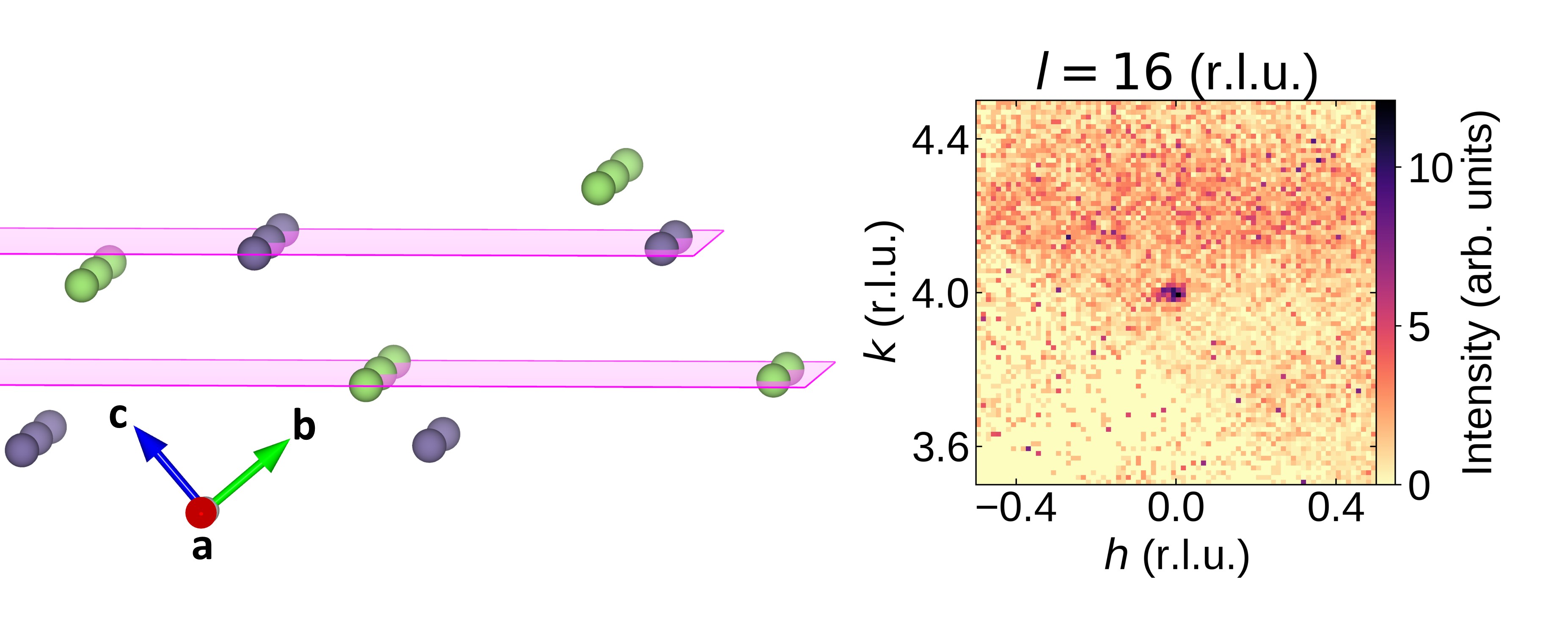}
        \end{minipage}
    \end{center}
    \caption[left]{
        Six examples, including the peak from Fig. \ref{fig:WhySg141_FoAreLower} in the main text, where $10^{-4}\leq\left|F_c\right|^2\leq1.21$ in Fig. \ref{fig:fitplots}(d) of the main text and the discrepancy between space groups $I4_1md$ and $I4_1/amd$ is most pronounced. In each case, the corresponding Bragg planes for space group $I4_1md$ can contain Ga with no Ge or vice versa.
	}
		\label{fig:6examplePeaksInTail}
\end{figure*}

In addition to Fig. \ref{fig:spectrumWideView} in the main text, Fig. \ref{fig:KeqPm1DataSlices} shows two other reciprocal space planes to demonstrate that the data for all 19 $\phi$ angles have merged well and that single Bragg peaks are seen at integer coordinates in reciprocal space. The only other features that are seen in Fig. \ref{fig:KeqPm1DataSlices} and Fig. \ref{fig:spectrumWideView} in the main text are powder scattering rings from the aluminum present in the sample holder, sample environment, and instrument windows.\\

\vspace{0.2cm}
\section{Quality of data where \bm{$10^{-4}\leq\left|F_c\right|^2\leq1.25$}}

Fig. \ref{fig:6examplePeaksInTail} shows 6 examples from the 1236 data points from figure \ref{fig:Fc109vsFc141} in the main text that give the strongest determination between the noncentrosymmetric $I4_1md$ symmetry, and centrosymmetric $I4_1/amd$ symmetry.

A defining feature of the 6 example peaks in Fig. \ref{fig:6examplePeaksInTail} is that the Bragg planes can have Ga atoms with no Ge atoms or vice versa.  Considering planes of pure Ga or Ge for $I4_1md$ symmetry, versus planes of mixed Ga and Ge for $I4_1/amd$ symmetry gives intuition as to why these structure factors for the $I4_1/amd$ symmetry would be significantly lower. 

It can be shown mathematically that the Bragg planes can contain Ga with no Ge or vice versa for each of the reflections shown in Fig. \ref{fig:6examplePeaksInTail}.
Let $\vec{b}_1$ be the atomic basis vector of a Ga atom in a Bragg plane with indices $h$, $k$, $l$.
Let $\vec{b}_2$ represent any of the atomic basis vectors of Ge.
Let $\vec{u}$ be the reciprocal lattice vector corresponding to the Bragg plane and $\vec{v}$ represent any Bravais lattice vector of the crystal.
The condition for a Bragg plane to contain Ga but no Ge is
\setlength{\abovedisplayskip}{3pt}
\setlength{\belowdisplayskip}{3pt}
\begin{equation}
    \label{PureGaCondit}
    \vec{u}\cdot\vec{b}_1\neq \vec{u}\cdot\left(\vec{b}_2+\vec{v}\right)
\end{equation}
for any $\vec{b}_2$ and any $\vec{v}$.
If $\vec{u}$ is expressed in terms of reciprocal lattice vectors and $\vec{v},\vec{b_1},\vec{b_2}$ are expressed in terms of direct lattice vectors, $\vec{u}\cdot\vec{v}$ can be any integer obtainable by integer combination of $h$, $k$, and $l$, so the condition for a Bragg plane to contain Ga but no Ge is
\begin{equation}
    \label{pureGaCondit_final}
    \vec{u}\cdot\left(\vec{b}_1-\vec{b}_2\right)\neq n
\end{equation}
where $n$ is, using Bezout's identity, any nonzero multiple of the greatest common divisor of $h$, $k$, and $l$.

Computations were carried out for each reflection in Fig. \ref{fig:Fc109vsFc141} in the main text to verify that corresponding Bragg planes contain Ga with no Ge or vice versa.

	\FloatBarrier
    \bibliographystyle{unsrtnat}
	\bibliography{si}